%% file: A-paper.tex
\documentclass[%
superscriptaddress,
12pt,
aps,
pre,
floatfix,
]{revtex4-1}
\usepackage{graphics}
\usepackage{lineno,hyperref}
\usepackage{graphicx}
\usepackage{url}
\usepackage{float}
\usepackage{latexsym}
\usepackage{amsmath,amssymb}
\usepackage{dcolumn}
\usepackage{epsfig}
\usepackage{textcomp}
\usepackage{multirow}
\usepackage{soul}
\usepackage{color}
\usepackage{booktabs}
\usepackage{Definitions}
\usepackage{enumitem}
\usepackage{subfig}
\usepackage{todo}
\usepackage{adjustbox}
\usepackage{amsfonts}
\usepackage{amsmath}
\usepackage{amssymb}
\usepackage{graphicx}
\usepackage{Definitions}
\usepackage{graphicx}
\usepackage{xcolor}
\usepackage{soul}
\usepackage{lineno}

\usepackage{hyperref} \hypersetup{colorlinks=true,citecolor=blue,linkcolor=blue,urlcolor=blue}
\usepackage{xcolor,soul}

\usepackage{fancyhdr}
\makeatletter \renewcommand\frontmatter@abstractwidth{\dimexpr\textwidth\relax} \makeatother 

\setstcolor{red} 
\bibpunct{[}{]}{,}{n}{}{} 

\newcommand{\our}{CrysXPP} 
\newcommand{\ourae}{CrysAE}
\newcommand{\xhdr}[1]{\vspace{1mm}\noindent{{\bf #1.}}}

\begin{document}
\title{\our{}: An Explainable Property Predictor for Crystalline Materials}
\author{Kishalay Das}
\affiliation{Indian Institute of Technology Kharagpur, Kharagpur, India}
\author{Bidisha Samanta}
\affiliation{Indian Institute of Technology Kharagpur, Kharagpur, India}
\author{Pawan Goyal}
\affiliation{Indian Institute of Technology Kharagpur, Kharagpur, India}
\author{Seung-Cheol Lee}
\email{leesc@kist.re.kr}
\affiliation{Indo Korea Science and Technology Center, Bangalore, India}
\author{Satadeep Bhattacharjee}
\email{s.bhattacharjee@ikst.res.in}
\affiliation{Indo Korea Science and Technology Center, Bangalore, India}
\author{Niloy Ganguly }
\email{niloy@cse.iitkgp.ac.in}
\affiliation{Indian Institute of Technology Kharagpur, Kharagpur, India }
\affiliation{Leibniz University of Hannover, Hannover, Germany}
\date{\today}
\begin{abstract}
We present a deep-learning framework, \our{}, to allow rapid and accurate prediction of electronic, magnetic, and elastic properties of a wide range of materials. \our{} lowers the need for large property tagged datasets by intelligently designing an autoencoder, \ourae{}. The important structural and chemical properties captured by \ourae{}  from the large amount of available crystal graphs data helped in achieving low prediction errors. Moreover, we design a feature selector that provides interpretability to the results obtained. Most notably, when given a small amount of experimental data, \our{} is consistently able to outperform conventional DFT.  A detailed ablation study establishes the importance of different design steps. We release the large pre-trained model \ourae{}. We believe by fine-tuning the model with a small amount of property-tagged data,  researchers  can achieve superior performance on various applications with a restricted data source.
\end{abstract}
\maketitle
\newpage
\input{BIntroduction}

\input{CDiscussion}
\input{DModel}
\vskip 0.5cm
\section{Data availability}
Most of the data used were from materials project. A small section of experimental data used as ground truth and as means to reduce the DFT bias will be available upon request to the corresponding authors. 
\vskip 0.5cm
\section{Code availability}
Source code of CrysAE and  \our{} is available in the following github repo:\\
\url{https://github.com/iitkgpaiforscience/Crysxpp.git}
\vskip 0.5cm
\section{competing interests}
The authors declare no competing interests.
\section{Author Contributions}
SB and NG conceived the idea, SCL and PG reviewed and helped to refine it. The model was developed by KD and BS, who also built the necessary codes. The first draft of the manuscript was written by KD, with contributions from BS. The paper was reviewed and approved by all authors, who all contributed to the final version of the manuscript.
\section{Corresponding authors}
Correspondence to Niloy Ganguly, Satadeep Bhattacharjee and Seung Cheol Lee.
\section{References}
\bibliography{main}
\bibliographystyle{naturemag}
\end{document}

%% file: BIntroduction.tex
\section{Introduction} 
In recent times several machine learning techniques ~\cite{seko2015prediction,xue2016accelerated,isayev2017universal,ghiringhelli2015big,isayev2015materials,xie2018crystal,sanyal2018mt,chen2019graph} have been proposed to enable fast and accurate prediction of different properties for crystalline materials, thus facilitating rapid screening of large material search spaces~\cite{meredig2014combinatorial,jha2018elemnet,choudhary2018machine}. The existing techniques either use handcrafted feature based descriptors ~\cite{seko2015prediction,xue2016accelerated,isayev2017universal,ghiringhelli2015big,isayev2015materials} or deep graph neural network (GNN) ~\cite{xie2018crystal,sanyal2018mt,chen2019graph,louis2020global,cheng2021geometric,banjade2021structure,jin2020hierarchical,qiao2020orbnet,ye2020symmetrical}  to generate a representation from the 3d conformation of crystal structures. Generating handcrafted features requires specific domain knowledge and human intervention, which make the methods inherently restricted. Deep learning methods, on the other hand, do not depend on careful feature curation and  can automatically learn the structure-property relationships of materials; thus making it an attractive candidate.\\
Graph neural network based approaches are getting  popular recently for their ability to encode graph information in an enriched representation space. Orbital-based GNNs \cite{qiao2020orbnet}\cite{ye2020symmetrical} use symmetry adapted atomic orbital features to predict different molecular properties. Though orbital-based GNNs predict molecular properties well, they are not an excellent choice for capturing complicated periodic structures such as crystals since they describe the nature of the electron distribution particularly close to atoms. On the other hand, motif-centric GNNs \cite{banjade2021structure} \cite{jin2020hierarchical} convert motif sub-structures of a crystal as a node and encode their inter connections for a large set of crystalline compounds using an unsupervised learning algorithm. Though they show improvements on property prediction tasks for metal oxides, their applicability is restricted as they ignore the atomic configuration inside the motif substructure which is also very important. \\
On a different departure, CGCNN \cite{xie2018crystal}, MTCGCNN \cite{sanyal2018mt} build a convolution neural network directly on a 2d crystal graph derived from 3d crystal structure. GATGNN \cite{louis2020global} incorporates the idea of graph attention network on crystal graphs to learn the importance of different bonds between the atoms whereas MEGNet \cite{chen2019graph} introduces global state attributes for quantitative structure-state property relationship prediction in materials. As this class of methods aims to capture the information of any crystal graph just from the connectivity and atomic features, we contribute in this promising direction. \\
Like any large deep neural network based models, GNN based architectures also introduce large number of trainable parameters. Hence, to estimate these parameters correctly for better accuracy, a huge amount of tagged training data is required  which is not always available for all the crystal properties. Hence developing a deep learning based model which can be trained on a small amount of tagged data would be extremely useful to infer varied properties of  crystal materials. Also as available experimental data for the various  properties are small and less diverse~\cite{kubaschewski1993materials,bracht1995properties,turns1995understanding}, these models are trained using data gathered from the DFT calculations~\cite{Materials_project,kirklin2015open,Wolverton}. As DFT data often differ from experimental ground truth due to its inefficiency in describing the many-body ground state, especially for properties such as  band gap~\cite{bandgap} or treatment of van der Waals interactions~\cite{van}, training with DFT only method may incorporate the inaccuracies of DFT in the  prediction. Moreover, in most of the cases, the existing property predictors are trained to predict a specific property. Hence, the generated descriptor or embeddings of any crystal are specific to a given property. It prevents them from sharing common structural information relevant to multiple properties. Though multi-task learning setup achieves information sharing across properties~\cite{sanyal2018mt}, it works well only for properties that are correlated with each other. 
 Last but not least, the existing  neural network based  methods~\cite{meredig2014combinatorial,ward2016general,jha2018elemnet,wu2018moleculenet,sanyal2018mt,xie2018crystal,louis2020global,chen2019graph,cheng2021geometric,banjade2021structure,jin2020hierarchical,qiao2020orbnet,ye2020symmetrical}  hardly provide any explanation for their results. The lack of interpretability and algorithmic transparency allows little use of them in the field of material science. Therefore it is necessary to explore and provide the reasons behind a prediction for any give property.
\\
In this paper, we propose an explainable deep property predictor ~\our{}. It is built upon  CrysAE, an auto-encoder based architecture  that is trained  with a large amount of  easily available crystal data, that is, property agnostic structural information of the crystal graph.  This leads to the deep encoding module capturing all the important structural and basic chemical information of the constituent atoms (nodes)  of the crystal graph. The learned information is leveraged to build the property predictor, CrysXPP, where the {\em knowledgeable} encoder helps to  produce high quality representation of  a candidate crystal. Consequently, the property predictor provides superior performance (better than all the competing baselines) even when  trained with a small amount of property-tagged data, thus largely mitigating the need for having a huge amount of  dataset tagged with a specific property. The structural information learned in the encoding model of an auto-encoder is robust and can  remove the error bias introduced by DFT by fine tuning the system with  a small amount of experimental data, whenever available. Further, we introduce a feature selector that helps to provide an explanation by highlighting the subset of the atomic features responsible for manifestation of  a chemical property of the given crystal.\\
Through extensive analysis of inorganic crystal data set across seven properties, we show that our method can achieve the lowest error compared to other alternative baselines; the improvement is particularly significant when only a small amount of tagged data is available for training. We have further shown that \our{} is effective towards removing error bias due to DFT tagged data by incorporating a  small amount of experimental data in the training set for both formation energy and band gap. Finally, with appropriate case studies, we show that the feature selection module can effectively provide explanations of the importance of different features towards prediction, which are in sync with the domain knowledge.

%% file: CDiscussion.tex
\section{Results and Discussions}
\label{sec:exp}
\subsection{Model Architecture}
\begin{figure*}[ht]
	\centering
	{\includegraphics[clip,width=\textwidth,height=60mm]{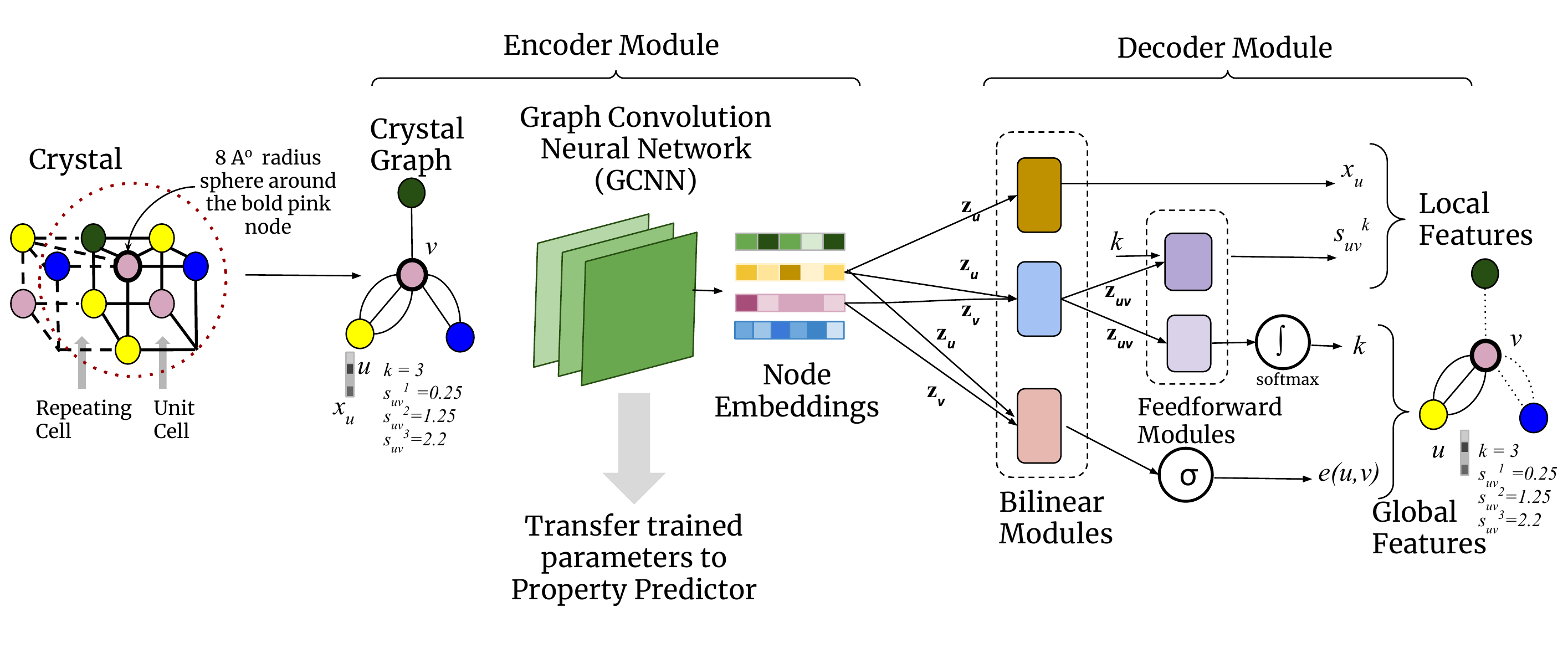}}
	\caption{The architecture of the Crystal Auto Encoder (CrysAE) module which comprises multilayer graph convolution network as the encoder and a set of decoding modules for reconstructing different local and global features.
	}
	\label{fig:Enc_architecture}
\end{figure*}

In this section we discuss in more detail the key technical contributions towards this goal followed by the training process and implementation details.
\begin{figure}[ht]
	\centering
	{\includegraphics[clip,width=\textwidth,height=60mm]{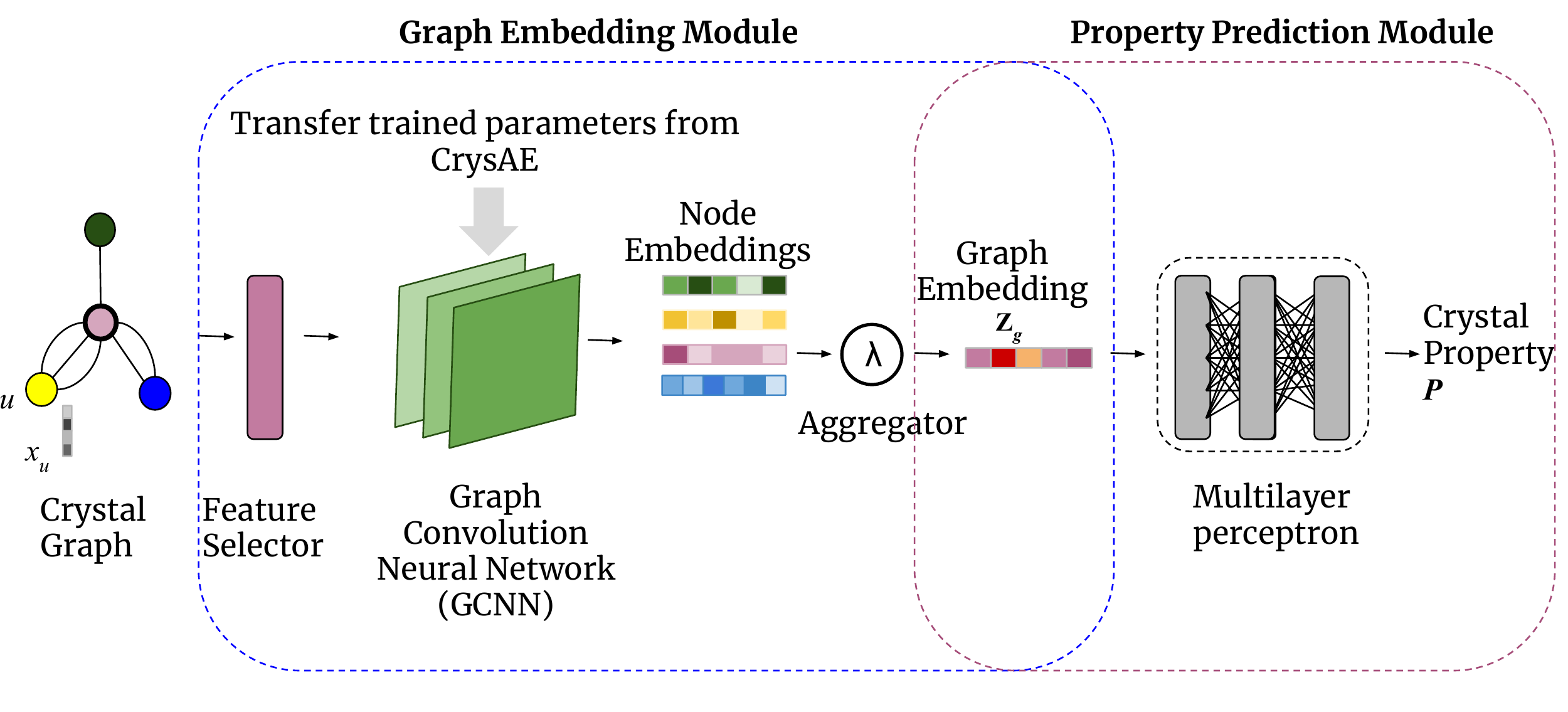}}
	\caption{The architecture of Crystal eXplainable Property Predictor (\our{}), comprises two building blocks (a). a multilayer graph convolution neural network (GCNN) as a graph embedding module and  (b). a multilayer perceptron as property prediction module. Given graph structure and node feature information, graph embedding module  produces an embedding corresponding to each graph. Property predictor  is a deep regressor module, which takes graph embedding as input and predicts the property value.}
	\label{fig:architecture}
\end{figure}\\
\xhdr{Overview}
We propose \textbf{Crys}tal e\textbf{X}plainable \textbf{P}roperty \textbf{P}redictor (\our{}), which realizes a crystalline material as a  graph structure (say \textit{$\Gcal$})  and  predicts the value of a property (eg. formation energy) given the crystal graph structure. As depicted in  Fig.\ref{fig:architecture}, \our{} comprises two building blocks (a). a property prediction module
and (b). a graph embedding module. In the graph embedding module we have a crystal graph  encoder based on  graph convolution neural network (GCNN) \cite{xie2018crystal}, which takes a crystal graph structure along with node and edge feature information as input and returns an embedding corresponding to each node as output. The weights of the node features (check Table \ref{tab:feature_desc}) are determined by a feature selector layer. We consider nine different atomic properties (Table \ref{tab:feature_desc}) as node features and the weights of those node features are determined by the  feature selector layer. Moreover,  the graph embedding module needs to capture the structural and chemical properties of the underlying crystal, hence one can use the huge amount of available crystal information (irrespective of the property) to train  the graph convolution network. 
For this at first we separately train the GCNN as a part (the encoder) of CrysAE (Fig.\ref{fig:Enc_architecture}); and the weights thereby obtained are used  as an initialization of the GCNN of \our{}. The structural information learned in the encoding model of CrysAE and duly transferred to the GCNN of \our{} makes \our{} more robust. \\
Our overall model architecture is essentially composed of the  following two modules:
\begin{itemize}
	\item \textbf{Auto encoder (CrysAE): } \\$q_{\bm{\theta}}: (\bm{\Vcal}, \bm{\Ecal}, \bm{\mathcal{X}}, \bm{\mathcal{F}}) \to \bm{\mathcal{Z}}$;
	$p_{\bm{\phi}}:  \bm{\mathcal{Z}} \to (\bm{\Vcal}, \bm{\Ecal}, \bm{\mathcal{X}}, \bm{\mathcal{F}})$
	\item \textbf{Property predictor (\our{}): }\\ $p_{\bm{\zeta}, \bm{\theta^\prime}, \bm{\psi}}: \bm{\mathcal{X}} \to_{\bm{\zeta}} \bm{\mathcal{X}}^\prime; (\bm{\Vcal}, \bm{\Ecal}, \bm{\mathcal{X}}^\prime, \bm{\mathcal{F}}) \to_{\bm{\theta^\prime}} \bm{\mathcal{Z}}; \bm{\mathcal{Z}} \to_{\bm{\psi}} \Pcal$
\end{itemize}
In the above characterization, $\bm{\theta}, \bm{\phi}, \bm{\zeta}, \bm{\theta}^\prime$ and $\bm{\psi}$ are the trainable parameters of the respective modules. Here $\bm{\theta}$ and $\bm{\phi}$ are the parameters for the encoder and decoder respectively of the CrysAE. $\bm{\zeta}$ is the trainable parameter of feature selector $\Scal$, $\bm{\theta}^\prime$ is the parameter of the encoder  and $\bm{\psi}$ is the parameter of the multi layer perceptron of \our{} model. We initialize $\bm{\theta}^\prime:=\bm{\theta}$ i.e, we first train the autoencoder and then the parameters of the encoder of CrysAE are transferred to the \our{}. 
\begin{table}[]
	\centering
	\small
	\setlength{\tabcolsep}{6pt}
	\scalebox{0.8}{
	\begin{tabular}{|c|c|}
		\hline
		\textbf{Features}  & \textbf{Feature Dimension}  \\
		\hline
		Group Number  & 18  \\
		\hline
		Period Number & 9\\
		\hline
		Electronegativity & 10 \\
		\hline
		Covalent Radius & 10 \\
		\hline
		Valence Electrons & 12\\
		\hline
		First Ionization Energy  & 10\\
		\hline
		Electron Affinity & 10 \\
		\hline
		Block & 4 \\
		\hline
		Atomic Volume& 10 \\
		\hline
	\end{tabular}
    }\centering
	\caption{Description of different properties used as atomic features and their dimensions. 
	}
	\label{tab:feature_desc}
\end{table}\\
\xhdr{Crystal Representation}
Our model realizes crystalline materials  as crystal graph structures $\bm{\mathcal{D}} = \{\Gcal_i =(\bm{\mathcal{V}}_i, \bm{\mathcal{E}}_i, \bm{\mathcal{X}}_i, \bm{\mathcal{F}}_i ) \}$ as proposed in~\cite{xie2018crystal}. 
Crystals have a repeating structure as depicted in Fig.\ref{fig:Enc_architecture} where a unit cell gets repeated across all the three dimensions. Hence, unlike simple graphs, the $\Gcal_i$ is an undirected weighted multi-graph where $\bm{\mathcal{V}}_i$ denotes a set of nodes (atoms) present in a unit cell of the crystal structure and $\bm{\mathcal{E}}_i = \{(u,v,k_{uv})\}$ denotes a multi-set of node pairs and the number of edges  between them.  $k_{uv}$ edges between a pair of nodes ($u,v$) indicate that $v$ is present in $k_{uv}$ repeating cells within $r$ radius from $u$ ($r$ is a hyper-parameter). 
$\bm{\mathcal{X}}_i$ represents node features i.e.  features that uniquely identify the chemical properties such as atomic volume, electron affinity, etc. of an atom as described in Table \ref{tab:feature_desc}. Lastly, $\bm{\mathcal{F}}_i$ corresponds to a muti-set of edge weights. We denote $\bm{\mathcal{F}}_i =\{\{s^k\}_{(u,v)} |  (u,v) \in \bm{\mathcal{E}}_i\}$ where $s^k$ denotes the $k^{th}$ bond length between the node pair $(u,v)$.  Between any pair of nodes, a maximum of $K$ edges are possible where $K$ is empirically determined. The bond length helps to specify the relative distance of an atom from its neighboring atoms. We use this graphical abstraction of a crystal as this can effectively embed the periodicity (indicated by the number of bonds) along with relative positioning for each atom in a simpler way, which otherwise was difficult to capture.  For easy reference, we drop the index of the notations. Next, we formally define the auto encoder (\ourae{}) and property predictor (\our{}).\\
\xhdr{Auto encoder (CrysAE)}
We build \textbf{Crys}tal \textbf{A}uto \textbf{E}ncoder (CrysAE) which composes of a simple encoder followed by an appropriate decoder to facilitate the overall training in order to learn  necessary information in the encoding mechanism. 
\\
\xhdr{Encoder}
We extend the crystal graph encoder proposed by Xie et.al.~\cite{xie2018crystal} to encode the chemical and structural information of a crystal graph $\Gcal$. Specifically, we encode $L$-hop neighbouring information of each node as: 
\small{
\begin{align}\label{eq:enc}
&\bm{h}_{(u,v)_k}^{l} = \bm{z}_u^{l} \oplus \bm{z}_v^{l} \oplus {s^k_{(u,v)}}\\
&\bm{z}_u^{l+1}  = \bm{z}_u^{l} + \sum\limits_{v,k} \sigma ( {\bm{h}^l_{(u,v)}}_k \bm{W}_c^{(l)} +  \bm{b}_c^{(l)}) \odot
g( {\bm{h}^l_{(u,v)}}_k \bm{W}_s^{(l)}+\bm{b}_s^{(l)}) \notag
\end{align}
}
where $\bm{z}_u^l$ denotes the embedding of node $u$ after $l$ hop neighbor information aggregation. The embedding of a node $u$ is initialized to a transformed node feature vector, i.e. it is a function of the atom $u$'s chemical features as $\bm{z}_u^0 := \bm{x}_u \bm{W}_x$ where $\bm{W}_x$ is the trainable parameter of the transformation network and $ \bm{x}_u $ is the input node feature vector. ${s^k_{(u,v)}} \in \bm{\mathcal{F}}_u $ represents the length of the $k^{th}$ edge between nodes $u$ and $v$. The $\oplus$ operator denotes concatenation and $\odot$ denotes element-wise multiplication. $\bm{W}_c^{(l)},\bm{W}_s^{(l)},\bm{b}_c^{(l)},\bm{b}_s^{(l)}$ are the convolution weight matrix, self weight matrix, convolution bias, self bias of $l^{th}$ hop convolution, respectively. $\sigma$ is a non-linear transformation function and it is used to generate a squeezed real value in [0,1] indicating the edge importance and $g$ is a feed forward network. 
After neighborhood aggregation we accumulate local information at each node which can be represented as $\bm{z}_u :=\bm{z}_u^L$. Subsequently we generate a graph level global information $\bm{\mathcal{Z}} = \{\bm{z}_1,..., \bm{z}_{|\bm{\mathcal{V}}|}\}$.
We do not aggregate the node embeddings further to prevent information loss in autoencoder.
We denote the set of trainable parameters for this encoder as $\bm{\theta}$ for future reference.
\\
\xhdr{Decoder}
We design an effective decoder that helps the encoder to transform the desired information in the representation vector space of $\bm{\mathcal{Z}}$. The decoder plays an inevitable role in learning the local and global structure as well as chemical features which are extremely useful. As mentioned earlier the global chemical features i.e. the crystal properties are a function of the local chemical environment and the overall conformation of the repeating  crystal cell structure; hence, we carefully design the decoder which can reconstruct two important features that induce the local chemical environment. They are (a) the node features i.e chemical properties of individual atoms and (b) local connectivity i.e the relative position of the nodes with respect to their neighbors. 
Precisely we reconstruct these information as below:
\begin{align}
&z_{uv} = \bm{z}_u^T \bm{W}_f \bm{z}_v + b_f \\
&\hat{s_{(u,v)}^k} = \begin{cases}
\gamma_s(z_{uv}\odot k) & \text{if } \gamma_s(.) > 0 \\
0 & \text{otherwise}
\end{cases}
\label{eq:l1}\\
&\bm{\mathcal{\hat X}}_{u} = \bm{W}_x^{T}\bm{z}_u + \bm{b}_x \label{eq:l2}
\end{align}
Eqs.~\ref{eq:l1} and~\ref{eq:l2} correspond to reconstructing the node property or atom's chemical property and a node's position relative to it's neighbors as we intend to achieve in (a) and (b) respectively. $z_{uv}$ is a combined transformed embedding of nodes $u$ and $v$ and $\gamma_s$ is a feed forward network which generates a real number corresponding to the length of the bonds. \\
Further we reconstruct the global structure i.e (c) the connectivity and periodicity of the crystal structures as below
\begin{align}
&(u,v) \sim p(e=(u,v)) = \sigma(\bm{z}_u^{T}\bm{W}_e\bm{z}_v + b_e ) \label{eq:g1}\\
&k_{(u,v)} = \arg \max_k \frac{e^{\gamma_k(z_{uv}, k)}}{\sum_k e^{\gamma_k(\bm{z}_{uv}, k)} } 
\label{eq:g2}
\end{align}
Here, $\bm{W}_e, b_e$ are trainable weight and bias associated with the bilinear edge reconstruction module, respectively. $\sigma$ is a squashing factor which provides a  value between $[0,1]$ denoting the edge probability. Similarly $\bm{W}_f, b_f$ are the trainable weight and bias parameters associated with the intermediate bi-linear transformation module, respectively. $\gamma_k$ represents a feed forward neural network that generates a $K$ length logit vector. We use a softmax to determine the exact number of edges present. Please note that though Eqs.~\ref{eq:g2},~\ref{eq:l1} correspond to global and local information respectively, they are heavily dependent upon each other, i.e the number of bonds and bond length both depend on the two end nodes information. Hence, we design a coupled embedding $z_{uv}$ which is shared by both the modules.
We denote the set of parameters in decoder as $\bm{\phi}$.\\
\xhdr{Training of auto-encoder} We learn the trainable parameters of both encoder and decoder by minimizing the reconstruction loss of different global and local structural and chemical features defined in Eqs.~\ref{eq:l1}-\ref{eq:g2}.
We minimize the cross-entropy loss of the predicted global features and node features along with mean squared loss of the edge weight or bond length in the  following objective: 
\begin{align}
&\EE_{\Gcal \sim \bm{\mathcal{D}}} -\sum_{(u,v) \in \Ecal} \big[\text{log}\ p(e=(u,v)) + \text{log} \ p(k_{(u,v)})\big]  \nonumber \\
& -\sum_{u \in \Vcal}\text{log}\ p(\bm{\mathcal{\hat X}}_{u}) + \sum_{(u,v) \in \Ecal} \sum_{k \in [1,\dots,K]} (s^k_{(u,v)} - \hat{s^k_{(u,v)}})^2 
\label{eq:aetrain}
\end{align}
where $p(.)$ denotes the probability of any event.
Thus by minimizing the reconstruction loss we not only fine tune parameters of decoder but efficiently train the encoder to generate a rich $\bm{\mathcal{Z}}$ which facilitates decoder operations.\\
\xhdr{Property predictor (\our{})}
Next, we design a property predictor specific to a property that can take the advantage of the structural information that is learned by the encoder as described above. We generate a graph level representation using the same graph encoder module as described in Eq.\ref{eq:enc}, thus in a way transferring the rich encoded knowledge to the property predictor. Next, we use a symmetric aggregation function to generate a single vector as graph representation $\bm{\mathcal{Z}}_g$. Thus the obtained representation of the graph is  invariant of the node orderings. Then the obtained representation is fed to a multilayer perceptron which predicts the value of the properties. More formally the property predictor can be characterized as:
\begin{align}
&\bm{\mathcal{Z}}_g  = \Lambda(\bm{z}_1\dots, \bm{z}_{|\Vcal|}) \label{eq:pool}\\
& \Pcal = \Mcal_{\bm{\psi}}(\bm{\mathcal{Z}}_g) \label{eq:prop}
\end{align}
Here, $\Lambda$ is the aggregation function which is symmetric. $\Mcal$ denotes a multilayer perceptron that has a trainable  parameter set $\bm{\psi}$. \\
\xhdr{Feature Selection}
The node features are first passed through a feature 
selector which is a trainable weight vector that selects a weighted subset of important node level features $\bm{\mathcal{X}}^\prime$ for a given property of interest $\Pcal$. $\bm{\mathcal{X}}^\prime$ forms input to the encoder.
\begin{align}
&\bm{\mathcal{X}}^\prime = \Scal_{\bm{\zeta}}(\bm{\mathcal{X}}) ;
\bm{\mathcal{Z}}_g = \Lambda(\text{Encoder}_{\bm{\theta^\prime}}(\bm{\mathcal{V}}, \bm{\mathcal{E}}, \bm{\mathcal{X}}^\prime, \bm{\mathcal{F}})) \nonumber 
\end{align}
In the above set of characteristic equations, $\Scal$ is the feature selector and $\bm{\zeta}$ is its trainable weight. We will show how the weights chosen by the  feature selection layer help us to explain the role of a node feature in the manifestation of a particular property (viz. formation energy) of a crystal. \\
\xhdr{Training of \our{}} We train the property predictor after the autoencoder. We initialize the trainable parameter  $\bm{\theta}^\prime:=\bm{\theta}$ where $\bm{\theta}$ is trained in the autoencoding module. Thus we first transfer the trained information such that  the property predictor benefits from the inductive bias already learned by training the autoencoder.\\ 
We use a LASSO~\cite{li2006lasso} regression to impose sparsity on the feature selector layer. Intuitively, if some atomic features ($\bm{\mathcal{X}}^\prime$) are crucial to predict a chemical property of the crystal,  the corresponding feature selector value  will be high and conversely, if some feature is not so important, the corresponding feature selector value will be negligible.  Hence, along with property prediction loss we also consider the LASSO regression loss as formally represented below:
\begin{align}\label{eq:proppred}
\min_{\bm{\zeta}, \bm{\theta^\prime}, \bm{\psi}}  (\hat{\Pcal}-\Pcal)^2 + \lambda_{1} * \vert{\bm{\zeta}} \vert_{L_1} 
\end{align}
where $\bm{\zeta}$ denotes the trainable parameters of feature selector $\Scal$ and  $\lambda_{1}$ is a hyper parameter which controls the degree of the regularization imposed. Before reporting the results, we  briefly discuss about the dataset and the baselines used for comparison.
\subsection{Dataset}
We have used the Materials Project database for our experiments which consists of $\sim$36,835 crystalline materials and is diverse in structure having materials with 87 different types of atoms, seven different lattice systems and 216 space groups. The unit cell of any crystal can have a maximum of 200 atoms. We consider nine properties for each atom which were used to construct the feature vector of each node \cite{xie2018crystal}. The details of the properties are given in Table~\ref{tab:feature_desc}. We convert them to categorical values if they are already not in that form. The dataset also provides DFT calculated target property values for the crystal structures. Experiments were done on a smaller training set than the original baseline papers.
\subsection{Comparison with similar baseline algorithms}
We compare the performance of \our{} with four state-of-the-art algorithms for crystal property prediction. These selected competing methods are  varied in terms of input data processing and working paradigms as described below:
\begin{enumerate}[label={(\alph*)}]
	\item \textbf{CGCNN}~\cite{xie2018crystal}: This  method generates crystal graphs from inorganic crystal materials and builds a graph convolution based supervised model for predicting various properties of the crystals.
	
	\item \textbf{MT-CGCNN}~\cite{sanyal2018mt} : This model uses the graph convolution based encoding as proposed in the previous model. Moreover, it incorporates multitask learning to jointly predict multiple properties of a single material.
	
	\item \textbf{MEGNET}~\cite{chen2019graph}: Here authors improved the CGCNN model further by introducing global state attributes including  temperature, pressure, entropy etc for quantitative structure-state-property relationship prediction in materials. Doing so they found that the crystal embeddings in MEGNet model encode periodic chemical trends. Further to address the issue of data limitation the 
	embeddings from a MEGNet model trained on formation energies is transferred and used to improve the accuracy of ML models for the band gap and elastic moduli.
	
	\item \textbf{ELEMNET}~\cite{jha2019enhancing}: This work does not specifically consider any structural properties of the crystal graph, rather it considers only the compositional atoms. It uses deep feed-forward networks to implicitly capture the effect of atoms on each other. It uses transfer learning to mitigate the error bias of DFT tagged data.
	
	\item \textbf{GATGNN}~\cite{louis2020global}: In this work authors have incorporated a graph neural network with multiple graph-attention layers (GAT) and a
    global attention layer, which can learn efficiently the importance of different complex bonds shared among the atoms within each atom’s local neighborhood.
	
\end{enumerate}
For all the baselines we have used the hyper parameters as mentioned in the original papers.
\subsection{Evaluation criteria}
We predict seven different properties of crystals in our experiments. Out of these, four are crystal state properties, namely, (a)  Formation Energy, (b) Band Gap, (c) Fermi Energy, (d) Magnetic Moment, and three are elastic properties, namely, (e) Bulk Moduli, (f) Shear Moduli, and (g) Poisson Ratio. All of these properties significantly depend on the details of  the crystal structure except Magnetic Moment which is more dependent on the atomic/node specifications as the magnetic moment arises from the unpaired \textit{d or f} electrons in an atom. Also the size of the moment depends on the local environments ~\cite{Lars,S1}. Moreover, we have very little DFT tagged data for Magnetic Moment and Band Gap.\\
We  focus on three different evaluation criteria as described below:
\begin{enumerate}
	\item How effective is the property predictor? Here we inspect the  performance of the  property predictor especially when it functions with a small amount of DFT tagged data.  
	\item How robust is the structural encoding? Here we investigate whether the structural encoding helps us to mitigate the noise introduced by DFT calculated properties.
	\item How effective is the explanation? We cross-validate the obtained explanation with domain knowledge.
\end{enumerate}
\begin{table*}[ht]
	\centering
	\small
	\setlength{\tabcolsep}{2.5pt}
	\scalebox{0.8}{
		\begin{tabular}{c c c c c c c c c}
		\toprule
		& \textbf{Property}& \textbf{Unit} & \textbf{CGCNN}  & \textbf{MTCGCNN } & \textbf{MEGNet} & \textbf{GATGNN} &  \textbf{Elemnet} & \textbf{\our{}} \\
		\hline
		\midrule
		\multirow{3}{*}{\rotatebox[origin=c]{90}{\shortstack{State \\ Properties}}}
		& Formation Energy   & eV/atom & 0.127 & 0.112 (0.147) & 0.142 & 0.164 & 0.098\textbf{*}  & \textbf{0.086} \\
		& Band Gap  & eV & 0.503 & 0.497 (0.518) & 0.498 & 0.489\textbf{*} & 0.491 & \textbf{0.467} \\ 
		& Fermi Energy  & eV & 0.528 & 0.503\textbf{*} (0.601) & 0.533 & 0.533 &  0.588 & \textbf{0.471} \\
		& Magnetic Moment  & $\mu_B$ & 1.21 & 1.16 (1.22) & 1.19 & 1.09 & \textbf{0.96} & 1.03\textbf{*}  \\
		\hline\hline
		\multirow{3}{*}{\rotatebox[origin=c]{90}{\shortstack{Elastic \\ Properties}}}
		& Bulk Moduli  & log(GPa) & 0.09 & 0.09 (0.09) & 0.105 &0.088\textbf{*} & 0.1057 & \textbf{0.08} \\
		& Shear Moduli  & log(GPa) & 0.125\textbf{*} & 0.120 (0.078) & 0.187 & 0.123 &  0.148 & \textbf{0.105}  \\
		& Poisson Ratio  &  - & 0.04 & 0.037\textbf{*} (0.039)& 0.041 & 0.039 &  0.039 & \textbf{0.035} \\
		\bottomrule
	\end{tabular}
    }
	\caption{Summary of the prediction performance (MAE) of different properties trained on 20\%  data and evaluated on $80\%$ of the data. The best performance is highlighted in bold and second best with \textbf{*}. We report MAE jointly training most correlated property (average on all property pairs) for MTCGCNN.}
	
	\label{tab:mt_mape}

\end{table*}
\subsection{Effectiveness of Property Predictor}
\begin{figure*}[h!]
	\centering
	\includegraphics[width=\columnwidth]{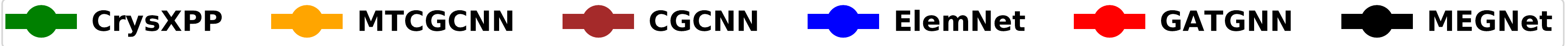}
	\\
	\vspace*{-1mm}
	\subfloat[Formation Energy]{
		\includegraphics[width=0.3\columnwidth, height=40mm]{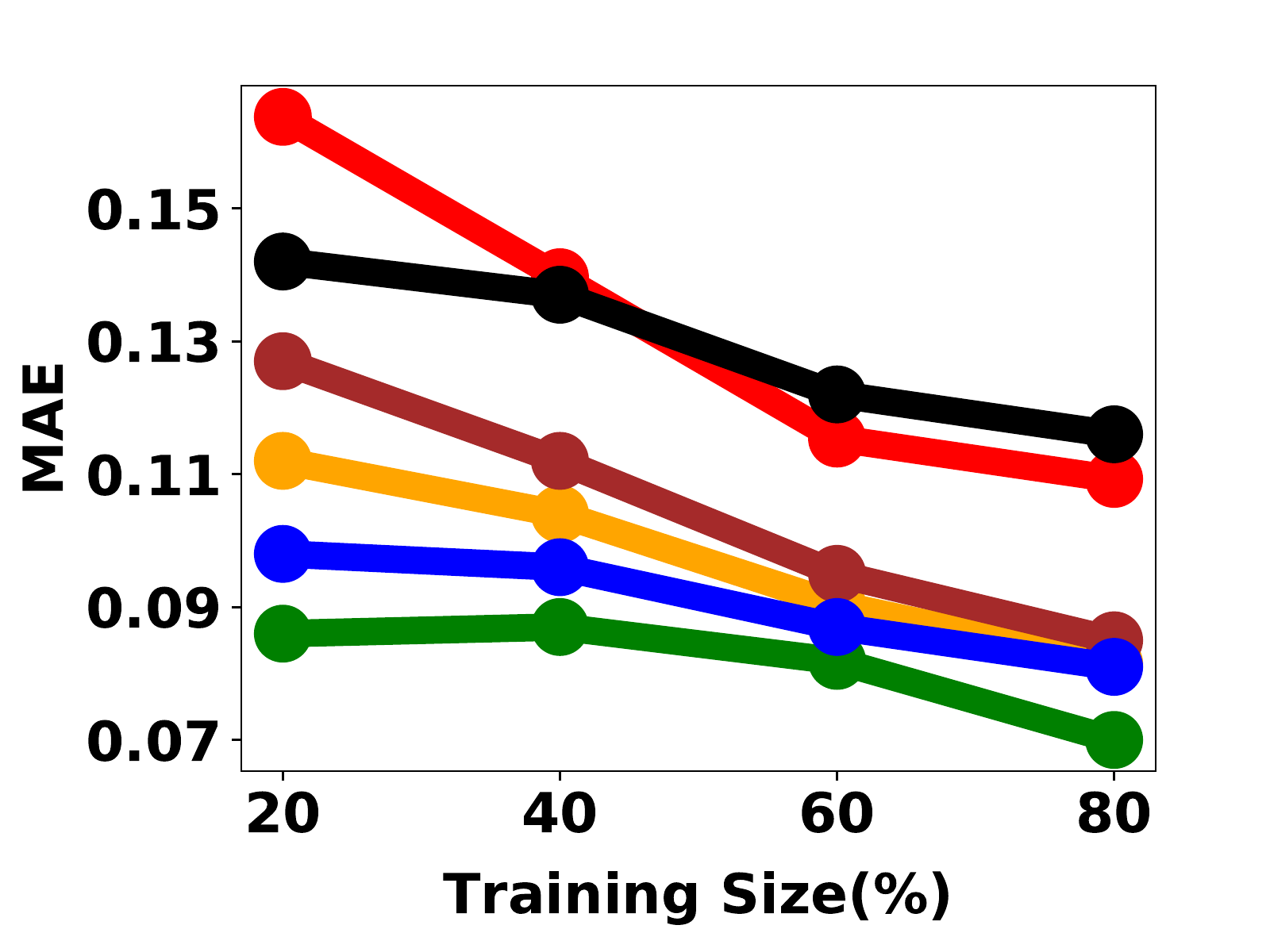}}
	\subfloat[Band Gap]{
		\includegraphics[width=0.3\columnwidth, height=40mm]{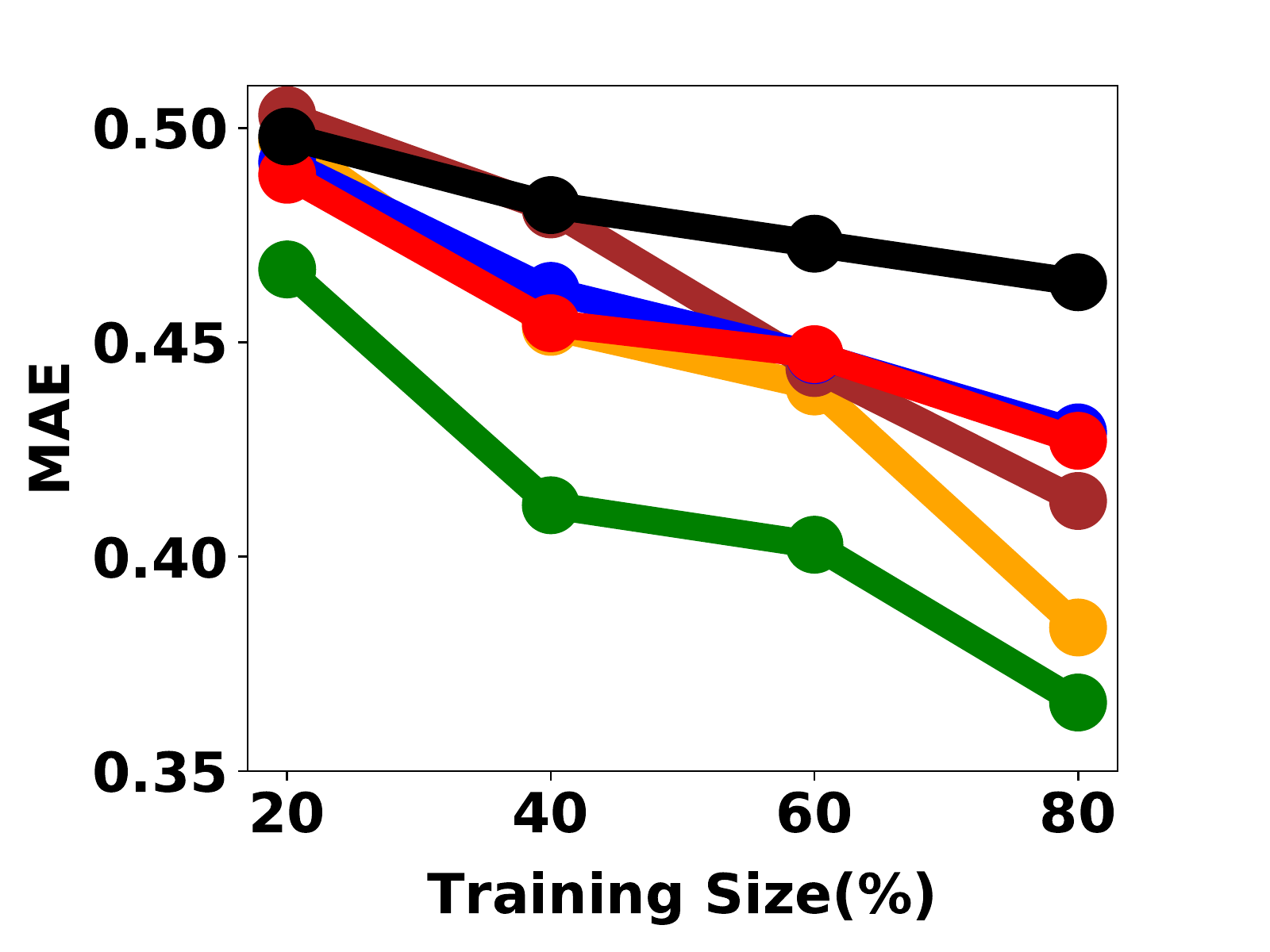}}
	\subfloat[Fermi Energy]{
		\includegraphics[width=0.3\columnwidth, height=40mm]{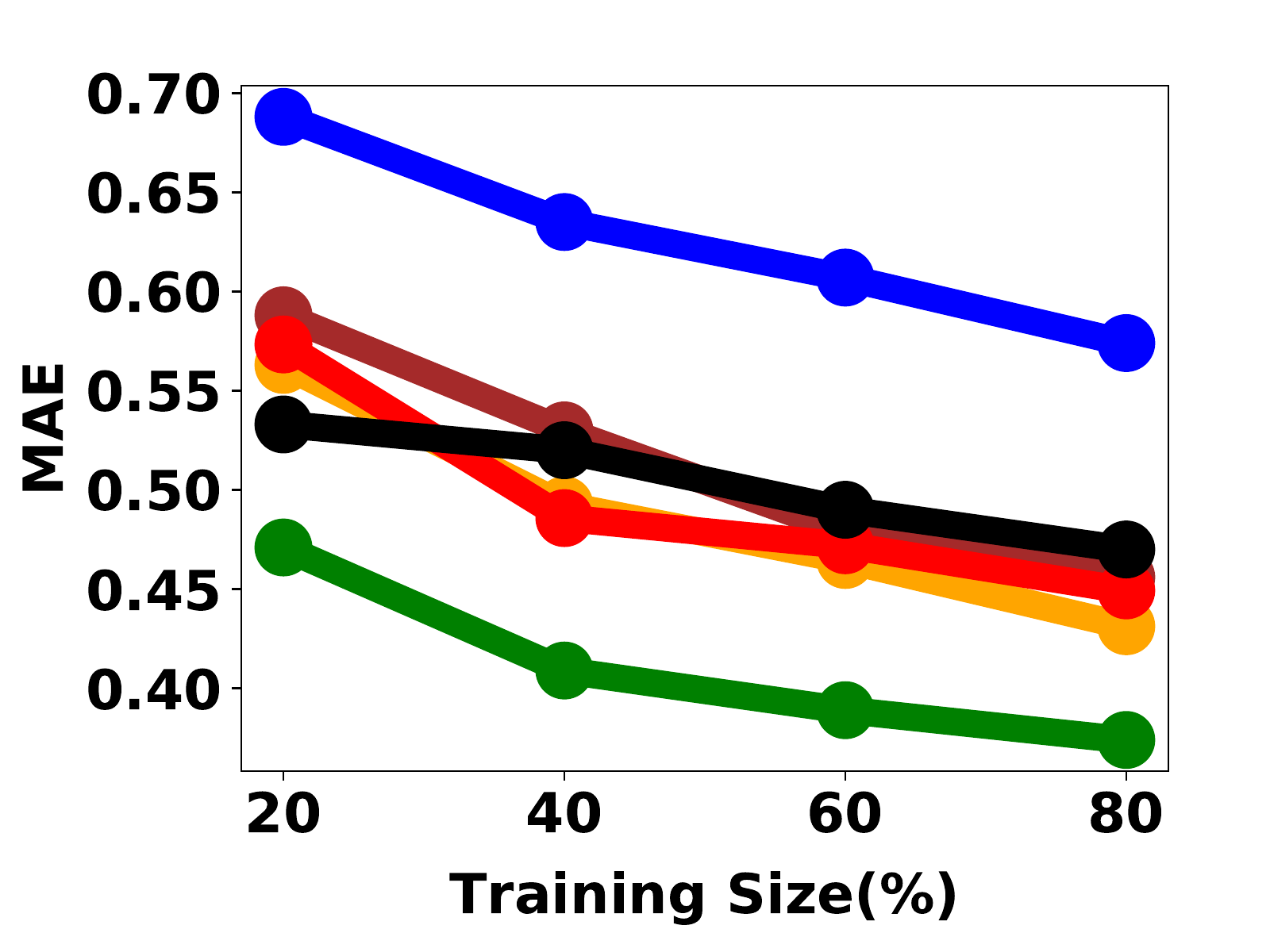}}\\
	\subfloat[Magnetic Moment]{
		\includegraphics[width=0.3\columnwidth, height=40mm]{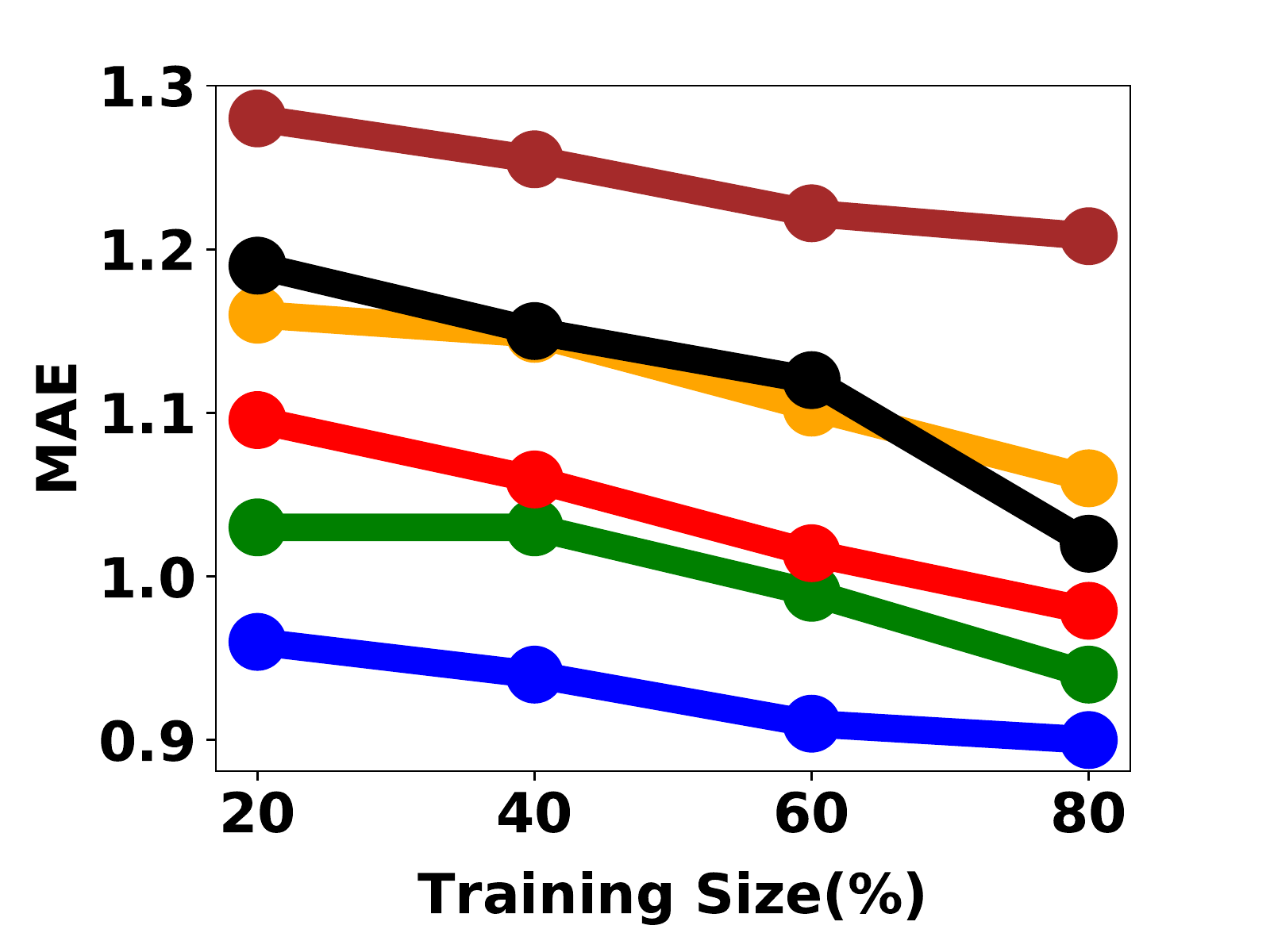}}
	\subfloat[Shear Moduli]{
		\includegraphics[width=0.3\columnwidth, height=40mm]{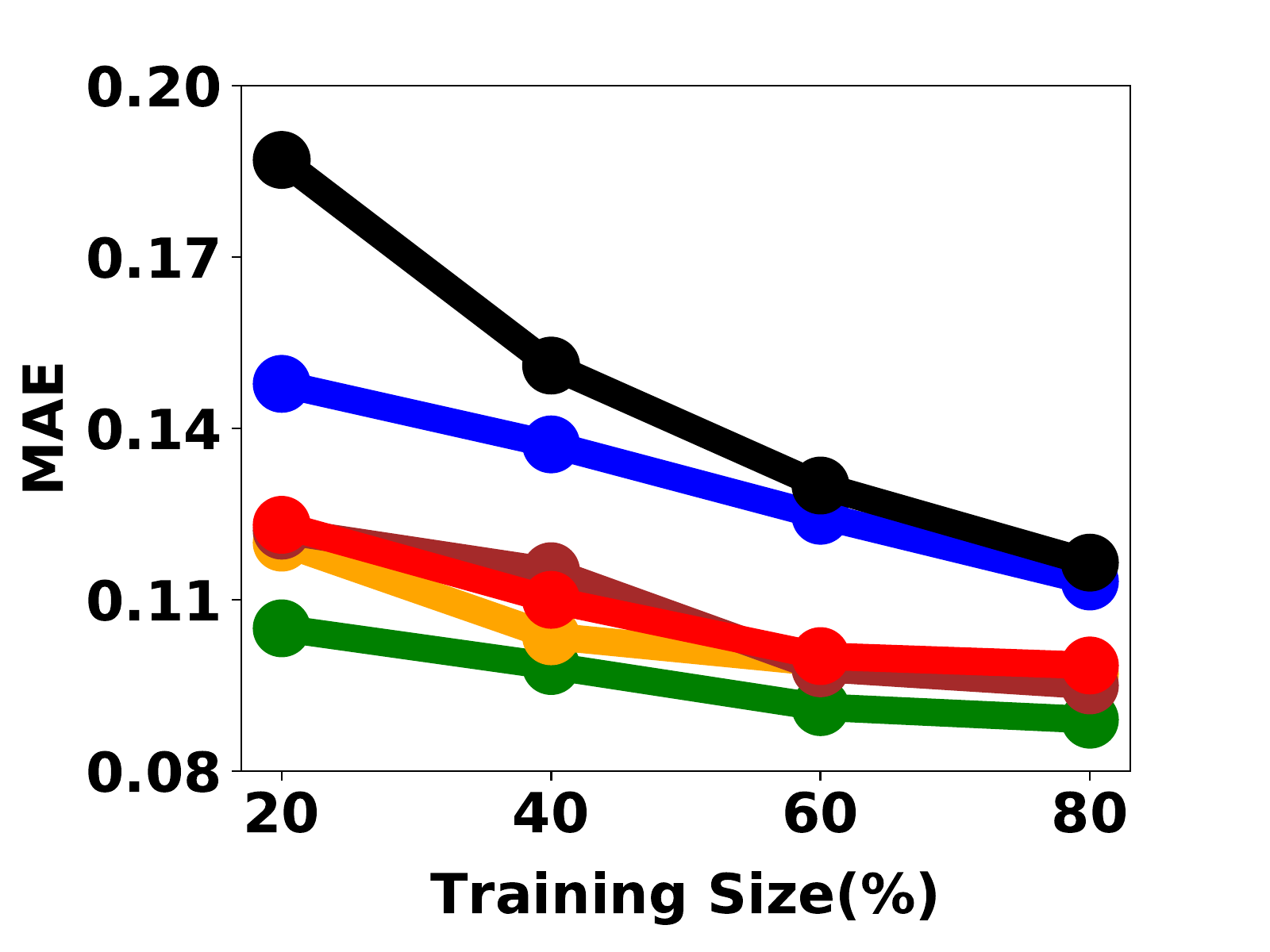}}
	\subfloat[Bulk Moduli]{
		\includegraphics[width=0.3\columnwidth, height=40mm]{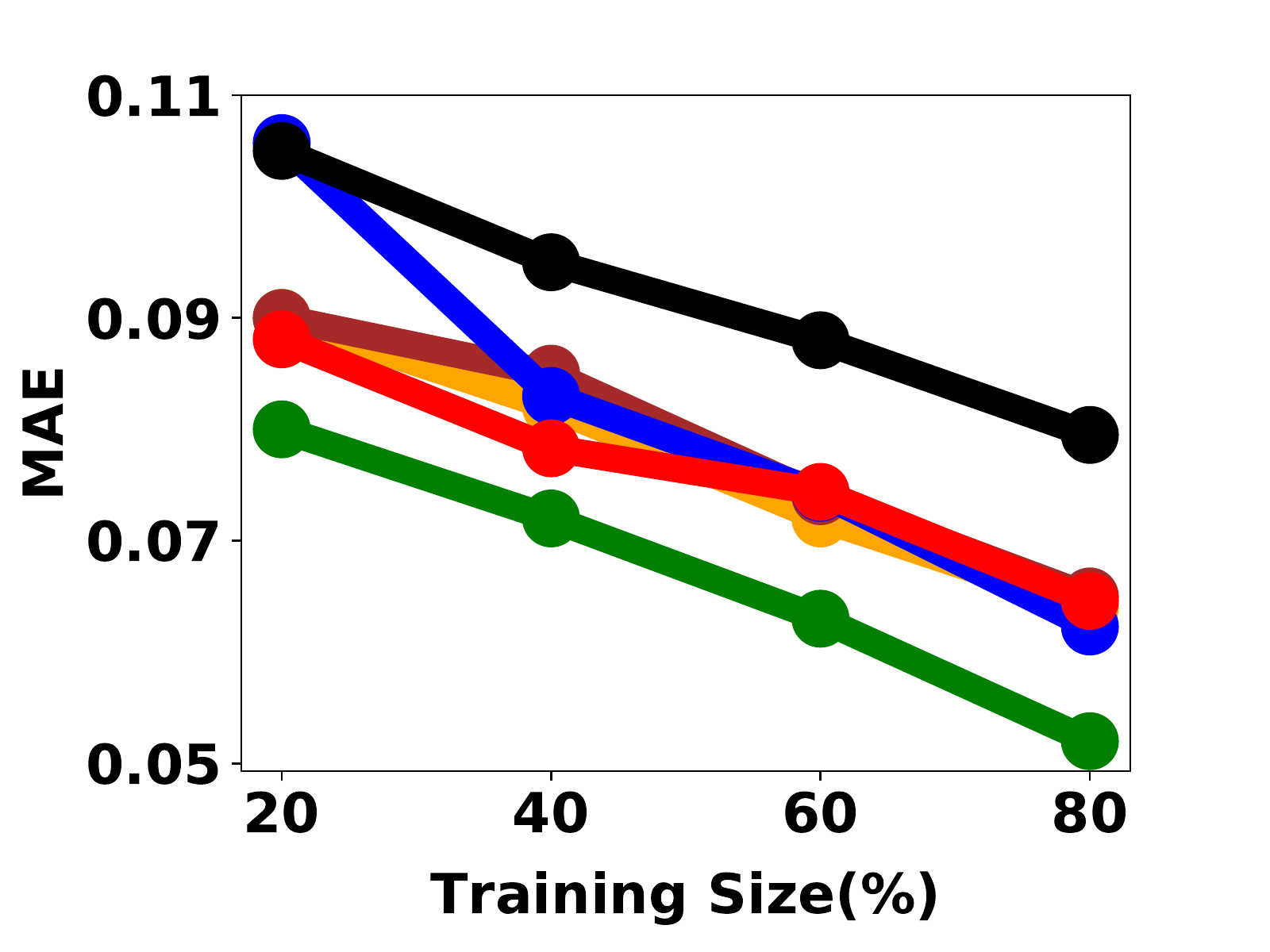}}\\
	\subfloat[Poisson Ratio]{
		\includegraphics[width=0.3\columnwidth, height=40mm]{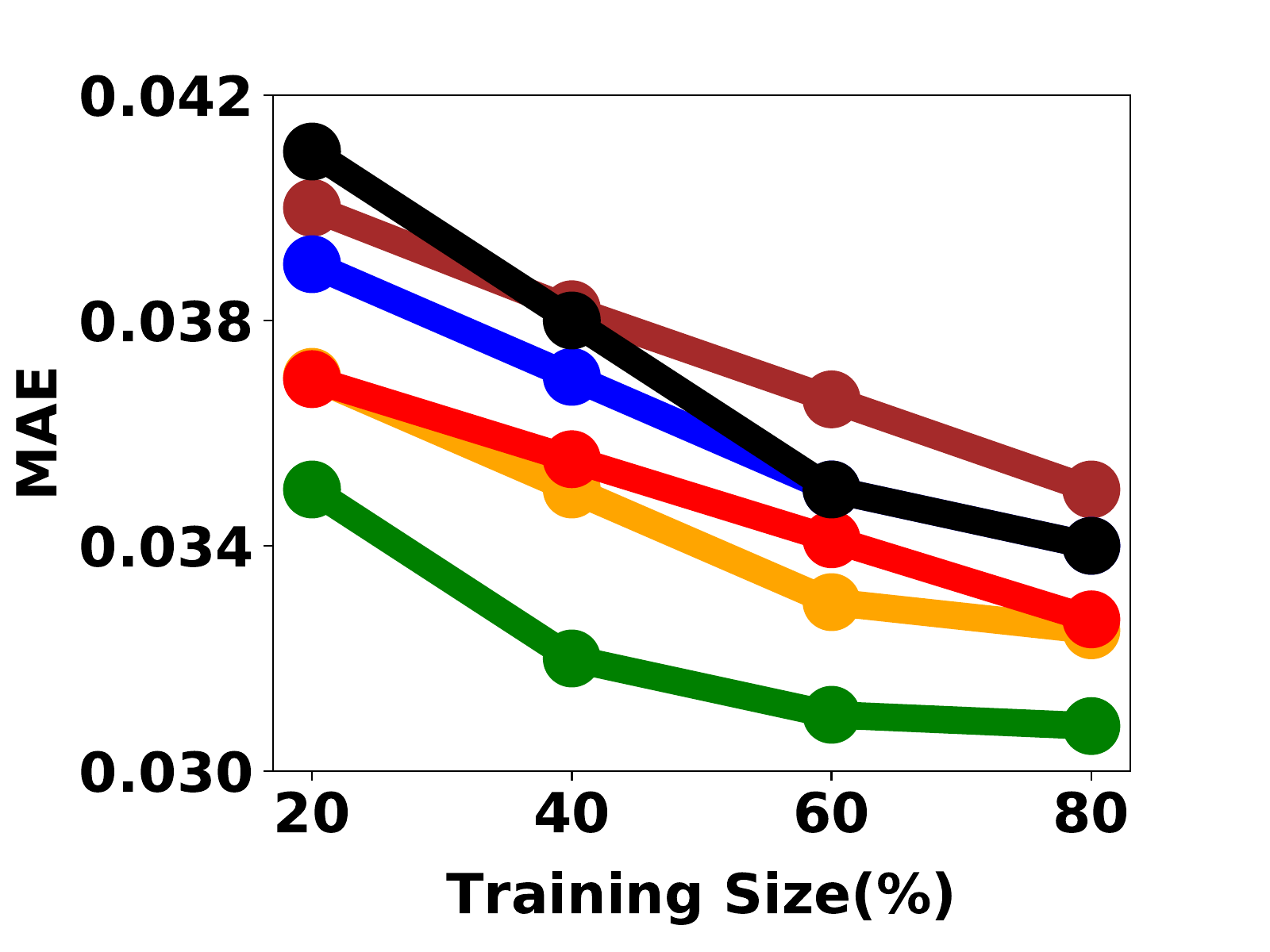}}
	
	\caption{Variation of MAE with the increase in training instances from 20\% to 80\%. \our{} outperforms all the baselines consistently.}
	\label{fig:results_on_more_train_data}
\end{figure*}

We first train the autoencoder with all untagged crystal graph present in the dataset, which captures all the structural information of the crystal graphs. Next for a given property of interest, we train the property predictor with 20\% of the available DFT tagged data and test on the rest. We report the 10 fold cross validation results.\\
\xhdr{Metric} We report Mean Absolute Error (MAE) to compare the performance of the participating methods. MAE is defined as $ \frac{1}{|\bm{\mathcal{D}}|}\sum_{\Gcal \in \bm{\mathcal{D}}}\big|{\Pcal_\Gcal - \hat \Pcal_\Gcal} \big|$, where $\Pcal_\Gcal$ is the property value calculated by DFT and $\hat \Pcal_\Gcal$ is the predicted value of a graph $\Gcal$.\\
\xhdr{Results}
In Table~\ref{tab:mt_mape} we report the MAE for \our{} as well as other alternatives on seven property values. We observe that \our{} outperforms every baseline across all the properties except magnetic moment. For MTCGCNN we report two values: the MAE obtained while jointly predicting the most correlated property, and the average MAE across all possible combinations (in bracket). It is interesting to note that its performance significantly degrades if the other property is not correlated with the current property of interest. A careful inspection reveals that for elastic properties, graph neural network based methods perform better than that of Elemnet. Elemnet only considers the composition of the crystal and ignores the global structural information, whereas these properties heavily depend on the crystal structure. In contrast, for Magnetic Moment the local information is important and hence, Elemnet performs the best and \our{} is the second best method. For the rest of the crystal state based properties, there is no consistent second best method. However, \our{} is a clear winner with a considerable margin which is due to the fact that the property predictor benefits from the structural knowledge transferred from the autoencoder.\\
\xhdr{Behaviour with increase in tagged data}
Further, we check the robustness of \our{}, by increasing  the percentage of tagged training data for property prediction. We  report the behavior of \our{} as well as other baselines in Fig.\ref{fig:results_on_more_train_data} for all the properties. We observe a monotonic decrease of MAE between predicted and  DFT calculated vales for most of the models where \our{} yields consistently smaller MAE and maintains the leadership position for all the properties expect Magnetic Moment.
This shows the robustness of our model to be able to perform consistently across a diverse set of properties with varied training instances. The MAE margin between \our{} and closest competitor (which is variable across properties), however, reduces  as training size increases. For Magnetic Moment, the local chemical information is more vital, hence ElemNet, which concentrate more on local chemical information, shows the best performance.
\subsection{ Removal of DFT error bias }
An important aspect of the prediction is that since we rely only on DFT data for training, we would be limited by the inaccuracies of DFT. In this section, we investigate with a system where we further fine tune the model with a small amount of available experimental data and  check whether the system can remove the   error propagated due to DFT. \\
\xhdr{Calculation setup}
We consider a  property predictor (as explained before) which has been trained with crystals whose particular property (e.g. Band Gap) values have been theoretically derived using DFT. We then fine tune the parameters with limited amount of experimental data; we perform it   for two different properties, namely, Band Gap and Formation Energy. For Formation Energy, we use 1,500 instances available at~\cite{kirklin2015open} and use different percentages of the data to fine-tune the model parameters. For Band Gap, we collect 20 experimental instances out of which we randomly pick 10 instances to fine-tune the parameters and report the prediction value for the rest ~\footnote{\url{http://matprop.ru}}. 
\begin{table}[]
    \centering
    \small
    \setlength{\tabcolsep}{3pt}
    \scalebox{0.75}{
    \begin{tabular}{l c c c c c c}
    \toprule
	\textbf{Experiment Settings} & \textbf{CGCNN} & \textbf{MTCGCNN} & \textbf{GATGNN} & \textbf{MEGNet} & \textbf{ElemNet} & \textbf{\our{}}\\
	\hline
	\midrule
	\textbf{\vtop{\hbox{\strut Train on 20\% DFT }\hbox{\strut Test on Full Experimental Data}}}  & 0.24 & 0.74 & 0.30 & 0.28 & \textbf{0.215}  & 0.22 \\
	\hline
	\textbf{\vtop{\hbox{\strut Train on 20\% DFT, 20 \% Experimental Data }\hbox{\strut Test on 80 \% Experimental Data}}} & 0.21 & 0.24 & 0.23 & 0.23 & 0.16\textbf{*} & \textbf{0.15} (0.206)  \\
	\hline
	\textbf{\vtop{\hbox{\strut Train on 80\% DFT, 20 \% Experimental Data }\hbox{\strut Test on 80 \% Experimental Data}}} & 0.16 & 0.22 & 0.19 & 0.18 & 0.1344\textbf{*} & \textbf{0.1319} (0.195)  \\
	\hline
	\textbf{\vtop{\hbox{\strut Train on 80\% DFT, 80 \% Experimental Data }\hbox{\strut Test on 20 \% Experimental Data}}} & 0.12 & 0.15 & 0.13 & 0.125 & 0.0905\textbf{*} & \textbf{0.0892} (0.174) \\
	\hline
    \bottomrule
    \end{tabular}
	}
    \caption{MAE of predicting experimental values after fine tuning different methods with different percentages of experimental data {for formation energy}. MAE of the experiment where we replace the experimental data with the same amount of DFT data to train \our{}, is provided in the bracket. The closest prediction is marked in bold and second best with \textbf{*}.
    }
    \label{tab:fe}
\end{table} 
\begin{table}[ht]
	\centering
	\small
	\setlength{\tabcolsep}{4pt}
	\scalebox{0.9}{
		\begin{tabular}{c c c c c c c c c c}
			\toprule
			\textbf{Materials} &  \textbf{Exp} & \textbf{DFT} & \textbf{\our{}} & \textbf{CGCNN} & \textbf{MTCGCNN} & \textbf{GATGNN} & \textbf{MEGNet} & \textbf{ElemNet}  \\
			\hline
			\midrule
			GaSb  & 0.72 & 0.36 & \textbf{0.77} (0.9\textbf{*}) & 1.01 (4.27)  & 0.26 (0.06) & 3.78(1.58)& 2.26(1.31) & 1.09 (0.33) \\
			\hline
			GaP  & 2.26 & 1.69 &  \textbf{2.10} (1.86) & 1.95\textbf{*} (1.49) & 0.73 (0.64) & 2.77(0.08)& 1.21(2.51) & 2.80 (1.29) \\
			\hline
			GaAs  & 1.42 & 0.18 & 1.54\textbf{*} (1.56) & 2.51 (\textbf{1.42}) & 1.83 (1.90) & 2.73(3.50)& 0.77(0.98) &0.83 (0.76)\\
			\hline
			InN  & 1.97 & 0.47 & \textbf{1.92} (1.85\textbf{*}) & 1.30 (2.90) & 1.77 (2.30) & 2.79(0.08) &1.33(2.16) &1.64 (1.43) \\
			\hline
			GaN  & 3.2 & 1.73 & 2.11 (1.47) & 3.51\textbf{*} (1.55) & 0.28 (0.16)&  2.27(0.56)& 1.59(2.66) &\textbf{3.69} (1.44)\\
			\hline
			NiO  & 4.3 & 2.214 & \textbf{2.45} (2.08) & 0.96 (1.12) & 0.08 (0.05)& 2.12(1.36) & 2.01(2.71) &2.31\textbf{*} (1.88)\\
			\hline
			Si  & 1.12 & 0.85 & \textbf{1.08} (0.95\textbf{*}) & 1.56 (1.64) & 0.39 (0.22) & 3.60(0.31)& 1.86(1.69) &0.33 (0.17)\\
			\hline
			ZnO  & 3.37 & 1.05 & 3.42\textbf{*} (2.1) & \textbf{3.32} (1.45) & 0.83 (0.56) & 2.74(1.36)& 2.09(2.53) &2.55 (2.01)\\
			\hline
           FeO  & 2.4 & 0 & \textbf{2.25} (1.72) & 2.16\textbf{*} (2.85) & 1.12 (0.96) & 1.92(1.02)& 2.93(2.81) &1.44 (1.27) \\
			\hline
			MnO  & 4 & 0.20 & \textbf{2.31} (1.81) & 1.51 (1.22) & 1.04 (0.77) & 2.44(2.35)& 1.73(2.11) &1.98\textbf{*} (1.44) \\
			\hline
			\bottomrule
		\end{tabular}
		
	}
	\caption{ Experiment (Exp) and predicted value for Band Gap for 10 crystals calculated by DFT and other machine learning models after fine-tuned by experimental data. The predicted value without fine tuning by experimental data is provided in the bracket.The closest prediction is marked in bold and the second best with  \textbf{*}. \our{} predicts closest to the ground truth after fine tuning with experimental data.}
	
	\label{tab:band_gap}
\end{table} \\ 
\xhdr{Results (Formation Energy)}
We report the mean absolute error (MAE) of Formation Energy in Table~\ref{tab:fe} achieved by different methods. 
The DFT prediction of the Formation Energy on the 1,500 crystals 
has an MAE of 0.21  with respect to experimental data and by training our model with DFT data we are performing close to the performance achieved by DFT. The results have a consistent trend for all the methods, whereby we observe that increasing the amount of training data, even if that is error-prone DFT data, helps in minimizing MAE. 
\our{} performs consistently better by a large margin than CGCNN and MTCGCNN, which takes the graph structure as an explicit input. However, it is interesting to observe that 
 ElemNet performs very close to \our{} as 
Formation Energy depends more on the composition than that on the explicit connection of atoms. Further, we conduct an experiment where we replace the experimental data with the same amount of DFT data to train our model. We then evaluate the performance of the model using experimental data as test data and find an inferior performance. We report the results in Table~\ref{tab:fe} (last column (in bracket)).
\\
\xhdr{Results (Band Gap)}
In Table~\ref{tab:band_gap} we report the experimental value of Band Gap for 10 test instances along with the predicted values by DFT and other machine learning methods. The error margin of DFT with the actual experimental values is quite high.
It is interesting to see that other than a few, DFT prediction is far from experimental data and in most of the cases, it is underestimating the experimental values.
 After fine-tuning DFT trained machine learning models with experimental data, the prediction becomes closer to the experimental value. However, \our{} performs closest to the experimental result in almost all the cases in comparison to other alternatives.  ElemNet, although second on the average when trained only on DFT (row 2 of Table \ref{tab:mt_mape}), cannot consistently maintain that position, whereas CGCNN performs better.  
 We have also provided results (in bracket) when we do not do any fine tuning. It can be seen even in such a scenario in many of the cases the performance is better than DFT. Further the power of \our{} in quickly mitigating the bias of DFT  when fine-tuned on minuscule data shows  the usefulness of modeling explicit structural information.
\begin{table}[]
    \centering
    \small
    \setlength{\tabcolsep}{4pt}
    \scalebox{0.9}{
    \begin{tabular}{c c c c c c} 
        \hline
        \multirow{2}{*}{\shortstack[t]{\textbf{Property}\\ \textbf{Name}}} &
        \multirow{2}{*}{\shortstack[t]{\textbf{Ablation }\\ \textbf{Settings}}} &
        \multicolumn{4}{c}{\textbf{Train-Test Split}}  \\ 
        \cline{3-6} & &  {\textbf{20\% - 80\%}} & {\textbf{40\% - 60\%}} & {\textbf{60\% - 40\%}} &{\textbf{80\% - 20\%}}\\
        \hline
        \hline
        
        \multirow{4}{*}{\shortstack[t]{\textbf{Formation}\\ \textbf{Energy}}}
        & Without Global + Local effect  & 0.124 & 0.113 & 0.092 & 0.086  \\
        \cline{2-6}
        & Global effect & 0.112 & 0.092 & 0.085 & 0.077 \\
        \cline{2-6}
        &  Local effect & \textbf{0.079} & \textbf{0.067} & \textbf{0.063} & \textbf{0.061} \\
        \cline{2-6}
        &  \our{} & 0.086 & 0.082 & 0.076 & 0.067  \\
       \cline{1-6}
       \cline{1-6}
       
       \multirow{4}{*}{\shortstack[t]{\textbf{Band}\\ \textbf{Gap}}}
       & Without Global + Local effect  & 0.502 & 0.482 & 0.452 & 0.408  \\
        \cline{2-6}
        & Global effect & 0.479 & 0.425 & 0.393 & 0.382 \\
        \cline{2-6}
        &  Local effect & 0.471 & 0.419 & 0.387 & 0.375 \\
        \cline{2-6}
        &  \our{} & \textbf{0.467} & \textbf{0.402} & \textbf{0.383} &  \textbf{0.366} \\
       \cline{1-6}
       \cline{1-6}
       
       \multirow{4}{*}{\shortstack[t]{\textbf{Fermi}\\ \textbf{Energy}}} 
       & Without Global + Local effect  & 0.513 & 0.481 & 0.477 & 0.443  \\
        \cline{2-6}
        & Global effect & 0.495 & 0.476 & 0.441 & 0.437 \\
        \cline{2-6}
        &  Local effect & 0.488 & 0.472 & 0.436 & 0.428 \\
        \cline{2-6}
        &  \our{} & \textbf{0.471} & \textbf{0.409} & \textbf{0.389} &  \textbf{0.374} \\
       \cline{1-6}
       \cline{1-6}
       
       \multirow{4}{*}{\shortstack[t]{\textbf{Magnetic}\\ \textbf{Moment}}}
       & Without Global + Local effect  & 1.082 & 1.038 & 1.023 & 1.027  \\
        \cline{2-6}
        & Global effect & 1.072 & 1.066 & 1.027 & 1.019 \\
        \cline{2-6}
        &  Local effect & 1.068 & 1.052 & 1.022 & 1.013 \\
        \cline{2-6}
        &  \our{} & \textbf{1.033} & \textbf{1.024} & \textbf{0.997} & \textbf{0.943}  \\
       \cline{1-6}
       \cline{1-6}
       
       \multirow{4}{*}{\shortstack[t]{\textbf{Bulk}\\ \textbf{Moduli}}} 
       & Without Global + Local effect  & 0.091 & 0.088 & 0.081 &  0.075 \\
        \cline{2-6}
        & Global effect & 0.088 & 0.081 & 0.075 & 0.068 \\
        \cline{2-6}
        &  Local effect & 0.087 & 0.077 & 0.072 & 0.063 \\
        \cline{2-6}
        &  \our{} & \textbf{0.080} & \textbf{0.072} & \textbf{0.063} & \textbf{0.052}  \\
       \cline{1-6}
       \cline{1-6}
       
       \multirow{4}{*}{\shortstack[t]{\textbf{Shear}\\ \textbf{Moduli}}} 
        & Without Global + Local effect  & 0.122 & 0.119 & 0.106 & 0.098  \\
        \cline{2-6}
        & Global effect & 0.119 & 0.108 & 0.099 & 0.093 \\
        \cline{2-6}
        &  Local effect & 0.117 & 0.102 & 0.097 & 0.091 \\
        \cline{2-6}
        &  \our{} &\textbf{ 0.105} & \textbf{0.098} & \textbf{0.091} & \textbf{0.089}  \\
       \cline{1-6}
       \cline{1-6}
       
       \multirow{4}{*}{\shortstack[t]{\textbf{Poisson}\\ \textbf{Ratio}}} 
       & Without Global + Local effect  & 0.039 & 0.035 & 0.033 &  0.032 \\
        \cline{2-6}
        & Global effect & 0.037 & 0.034 & 0.032 & 0.031 \\
        \cline{2-6}
        &  Local effect & 0.037 & 0.033 & 0.031 & 0.031 \\
        \cline{2-6}
        
        &  \our{} & \textbf{0.035} & \textbf{0.032} & \textbf{0.031} & \textbf{0.030}  \\
       \cline{1-6}
       \hline
    \end{tabular}
	}
    \caption{Summary of experiments of ablation study on importance of different reconstruction loss components on \ourae{} training and eventually its effect on \our{} (MAE for property prediction).}
    \label{tab:ablation_study}
\end{table}

\begin{table}[]
    \centering
    \small
    \setlength{\tabcolsep}{4pt}
    \scalebox{0.9}{
    \begin{tabular}{c c c c c c} 
        \hline
        \multirow{2}{*}{\shortstack[t]{\textbf{Property}\\ \textbf{Name}}} &
        \multirow{2}{*}{\shortstack[t]{\textbf{Ablation }\\ \textbf{Settings}}} &
        \multicolumn{4}{c}{\textbf{Train-Test Split}}  \\ 
        \cline{3-6} & &  {\textbf{20\% - 80\%}} & {\textbf{40\% - 60\%}} & {\textbf{60\% - 40\%}} &{\textbf{80\% - 20\%}}\\
        \hline
        \hline
        
        \multirow{2}{*}{\shortstack[t]{\textbf{Formation}\\ \textbf{Energy}}}
        & Without $L_{1}$ regularizer & 0.092 & 0.089 & 0.083 & 0.0731  \\
        \cline{2-6}
        &  With $L_{1}$ regularizer & \textbf{0.086} & \textbf{0.082} & \textbf{0.076} & \textbf{0.067}  \\
       \cline{1-6}
       \cline{1-6}
       
       \multirow{2}{*}{\shortstack[t]{\textbf{Band}\\ \textbf{Gap}}}
        & Without $L_{1}$ regularizer  & 0.476 & 0.417 & 0.391 & 0.374  \\
        \cline{2-6}
        & With $L_{1}$ regularizer & \textbf{0.467} & \textbf{0.402} & \textbf{0.383} &  \textbf{0.366} \\
       \cline{1-6}
       \cline{1-6}
       
       \multirow{2}{*}{\shortstack[t]{\textbf{Fermi}\\ \textbf{Energy}}} 
        & Without $L_{1}$ regularizer  & 0.502 & 0.441 & 0.415 & 0.394  \\
        \cline{2-6}
        &  With $L_{1}$ regularizer & \textbf{0.471} & \textbf{0.409} & \textbf{0.389} &  \textbf{0.374} \\
       \cline{1-6}
       \cline{1-6}
       
       \multirow{2}{*}{\shortstack[t]{\textbf{Magnetic}\\ \textbf{Moment}}}
        & Without $L_{1}$ regularizer  & 1.094 & 1.046 & 1.028 & 1.013  \\
        \cline{2-6}
        & With $L_{1}$ regularizer & \textbf{1.033} & \textbf{1.024} & \textbf{0.997} & \textbf{0.943}  \\
       \cline{1-6}
       \cline{1-6}
       
       \multirow{2}{*}{\shortstack[t]{\textbf{Bulk}\\ \textbf{Moduli}}} 
        & Without $L_{1}$ regularizer  & 0.093 & 0.082 &  0.067 & 0.061  \\
        \cline{2-6}
        &  With $L_{1}$ regularizer & \textbf{0.080} & \textbf{0.072} & \textbf{0.063} & \textbf{0.052}  \\
       \cline{1-6}
       \cline{1-6}
       
       \multirow{2}{*}{\shortstack[t]{\textbf{Shear}\\ \textbf{Moduli}}}
        & Without $L_{1}$ regularizer  & 0.128 & 0.115 & 0.099 & 0.095  \\
        \cline{2-6}
        &  With $L_{1}$ regularizer &\textbf{ 0.105} & \textbf{0.098} & \textbf{0.091} & \textbf{0.089}  \\
       \cline{1-6}
       \cline{1-6}
       
       \multirow{2}{*}{\shortstack[t]{\textbf{Poisson}\\ \textbf{Ratio}}}
        & Without $L_{1}$ regularizer  & 0.038 & 0.033 & 0.033 & 0.032  \\
        \cline{2-6}
        &  With $L_{1}$ regularizer & \textbf{0.035} & \textbf{0.032} & \textbf{0.031} & \textbf{0.030}  \\
       \cline{1-6}
       \hline
    \end{tabular}
	}
    \caption{Summary of experiments of ablation study on sparse  feature selection using $L_1$ regularizer, performed on different train test splits across different properties (MAE).}
    \label{tab:ablation_study_fs}
\end{table} 

 \begin{table}[]
    \centering
    \small
    \setlength{\tabcolsep}{4pt}
    \scalebox{0.9}{
    \begin{tabular}{c c c c c c} 
        \hline
        \multirow{2}{*}{\shortstack[t]{\textbf{Property}\\ \textbf{Name}}} &
        \multirow{2}{*}{\shortstack[t]{\textbf{Ablation }\\ \textbf{Settings}}} &
        \multicolumn{4}{c}{\textbf{Train-Test Split}}  \\ 
        \cline{3-6} & &  {\textbf{20\% - 80\%}} & {\textbf{40\% - 60\%}} & {\textbf{60\% - 40\%}} &{\textbf{80\% - 20\%}}\\
        \hline
        \hline
        
        \multirow{2}{*}{\shortstack[t]{\textbf{Formation}\\ \textbf{Energy}}}
        & GCN Encoder  & 0.187 & 0.162 & 0.138 & 0.125  \\
        \cline{2-6}
        &  CGCNN Encoder & \textbf{0.086} & \textbf{0.082} & \textbf{0.076} & \textbf{0.067}  \\
       \cline{1-6}
       \cline{1-6}
       
       \multirow{2}{*}{\shortstack[t]{\textbf{Band}\\ \textbf{Gap}}}
        & GCN Encoder   & 0.613 & 0.515 & 0.493 & 0.476  \\
        \cline{2-6}
        &  CGCNN Encoder & \textbf{0.467} & \textbf{0.402} & \textbf{0.383} &  \textbf{0.366} \\
       \cline{1-6}
       \cline{1-6}
       
       \multirow{2}{*}{\shortstack[t]{\textbf{Fermi}\\ \textbf{Energy}}} 
        & GCN Encoder   & 0.513 & 0.489 & 0.450 & 0.436  \\
        \cline{2-6}
        &  CGCNN Encoder & \textbf{0.471} & \textbf{0.409} & \textbf{0.389} &  \textbf{0.374} \\
       \cline{1-6}
       \cline{1-6}
       
       \multirow{2}{*}{\shortstack[t]{\textbf{Magnetic}\\ \textbf{Moment}}}
        & GCN Encoder   & 1.204 & 1.113 & 1.080 & 1.041  \\
        \cline{2-6}
        &  CGCNN Encoder & \textbf{1.033} & \textbf{1.024} & \textbf{0.997} & \textbf{0.943}  \\
       \cline{1-6}
       \cline{1-6}
       
       \multirow{2}{*}{\shortstack[t]{\textbf{Bulk}\\ \textbf{Moduli}}} 
        & GCN Encoder   & 0.171 & 0.133 &  0.114 & 0.101  \\
        \cline{2-6}
        &  CGCNN Encoder & \textbf{0.080} & \textbf{0.072} & \textbf{0.063} & \textbf{0.052}  \\
       \cline{1-6}
       \cline{1-6}
       
       \multirow{2}{*}{\shortstack[t]{\textbf{Shear}\\ \textbf{Moduli}}}
        & GCN Encoder   & 0.183 & 0.172 & 0.168 & 0.141 \\
        \cline{2-6}
        &  CGCNN Encoder &\textbf{ 0.105} & \textbf{0.098} & \textbf{0.091} & \textbf{0.089}  \\
       \cline{1-6}
       \cline{1-6}
       
       \multirow{2}{*}{\shortstack[t]{\textbf{Poisson}\\ \textbf{Ratio}}}
        & GCN Encoder   & 0.041 & 0.037 & 0.036 & 0.034  \\
        \cline{2-6}
        &  CGCNN Encoder & \textbf{0.035} & \textbf{0.032} & \textbf{0.031} & \textbf{0.030}  \\
       \cline{1-6}
       \hline
    \end{tabular}
	}
    \caption{Summary of experiments (MAE) of ablation study on effect of GCN as graph encoder in \ourae{} and \our{}.}
    \label{tab:ablation_study_gnn}
\end{table}
\subsection{Ablation Studies}
We demonstrate the effectiveness of architectural choices and training strategies for \our{}, by  designing the following set of ablation studies:
\begin{enumerate}
	\item The importance of explicitly capturing global and local features and understanding their effect on property prediction
	\item The impact of sparse feature selection on property prediction, and 
	\item The choice of GNN models in the autoencoder \ourae{}
\end{enumerate}
In the following subsections we will thoroughly discuss these.
\\
\xhdr{Importance of local and global feature understanding}
Here we investigate the importance of different reconstruction loss components on \ourae{} training and eventually its effect on property prediction.
 (a). Without Global + Local effect : In this scenario we do not train \ourae{} and only train \our{}.
(b). Global effect : We train \ourae{} by minimizing the reconstruction loss of only global features and ignoring local feature losses.
(c). Local effect: Here we focus only minimizing local feature losses.
We report the performance of the model (MAE) in Table \ref{tab:ablation_study} on different train test splits across different properties. We observe that the performance of the model in the setting  Without Global + Local effect, is the worst. We also notice that for all the properties, Local effect individually leads to better performance than  Global effect, except for the Poisson ratio where the effect is similar for both the cases. However, it is found that the impact of local and global effect are somewhat complementary, hence simultaneous reconstruction of 
global and local features (\our{}) results in the best performance. The only exception is  formation energy where 
addition of global feature leads to performance deterioration..
\\
\xhdr{Impact of sparse feature selection} We perform an ablation study  to analyze the impact of sparse feature selection on property prediction.
This is done by removing the $L_{1}$ regularizer term from \our{} loss function in Eq.\ref{eq:proppred}. We evaluate the performance of the model and report the results (MAE) in Table \ref{tab:ablation_study_fs}. We observe discernible improvement due to the introduction of sparse feature selection using $L_{1}$ regularizer.

\xhdr{Effect of other GNN variants as graph encoder} To explore the effectiveness of other GNN variants as graph encoders, we conduct an experiment where we replace the CGCNN encoder with one of the popular GNN variants:  GCN \cite{kipf2017semi} encoder and evaluate the performance of the model. GCN only considers the graph structural information and atom features to learn the graph representation and unlike CGCNN, it does not consider the individual edges weights in the multi-graph representing a crystal. We report the results of the model performance (MAE) in Table \ref{tab:ablation_study_gnn}. We observe that the model performance degrades when trained with GCN. The edge weight calculation, which is a major contribution of CGCNN, is extremely helpful to capture the local structure of the crystal.

\subsection{Explanation through feature selection}
We have introduced a feature selector that is trained along with the property prediction parameters with available tagged data. The feature selector helps to select the subset of the atomic features contributing to the chemical properties of the crystal which makes the model explainable by design. To demonstrate the effectiveness of the feature selector, we have selected few case studies and provide the feature explanation for formation energy, band gap and magnetic moment.
\begin{figure*}[h!]
	\centering
	\includegraphics[width=0.80\linewidth, height=48mm]{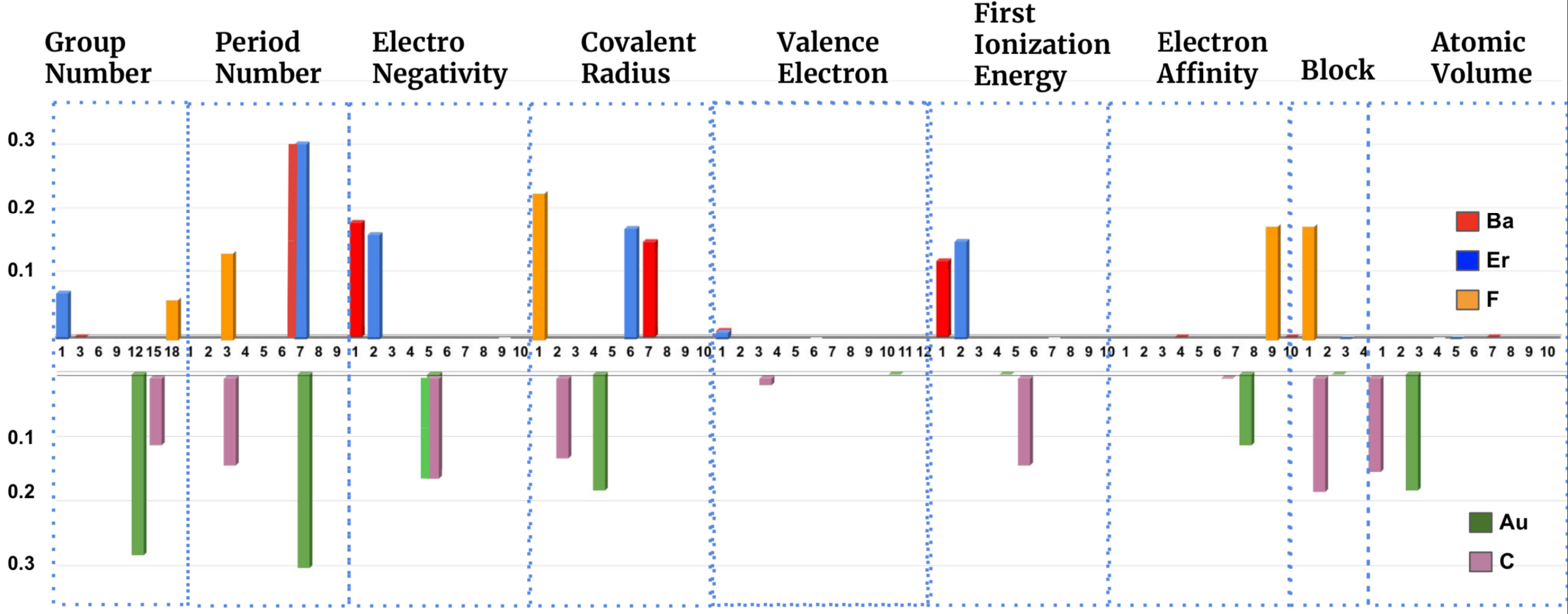}
	\caption{Feature selector values corresponding to atom features after trained on Formation Energy tagged data. The top bar chart represents the feature weights of $BaEr_2F_8$ 
	and the below one represents the feature weights of AuC.  
	}
	\label{fig:feature_mask_cs}
\end{figure*}\\
\xhdr{Formation Energy}
We here report case studies corresponding to two crystals $BaEr_2F_8$ and AuC,  illustrating the important role of feature selector in providing explanation. 
We report the feature selector values corresponding to categorical atomic properties after being trained on Formation Energy tagged data in Fig.\ref{fig:feature_mask_cs}.  The bars represent the weights assigned by the feature selector on the categorical values of atomic features and different colors indicate different atoms. Higher  category denotes higher value of the feature. Fig.\ref{fig:feature_mask_cs} depicts the importance of the atomic features in two extreme cases. One is $BaEr_2F_8$, whose Formation Energy is predicted as -4.41 eV/atom indicating its stability while the other is AuC with predicted Formation Energy 2.2 eV/atom  denoting the material is quite unstable. In both cases we see that Period Number is the most important atomic feature as it has maximum weight. Period and Group Numbers provide the
information to distinguish each element. 
As the Group Numbers and the number of
Valence Electrons are closely related, 
we see that the feature selector only selected the former thus avoiding duplicity.
Electronegativity and Covalent Radius both are another two important features (with non-zero weight) which is evident from the figure.  
Non-zero difference in Electronegativity of atoms indicates
stability in structure.
Both Au and C
have the same Electronegativity (category value 5), and feature selector gives same weight to it, as a result the difference of Electronegativity is zero in the case of unstable AuC. 
While
for the case of stable $BaEr_2F_8$, the feature selector provides different non-zero weights to smaller Electronegative elements and zero weight
to the largest Electronegative atom F.
The Covalent Radius determines the extent of overlap of electron densities of constituents, therefore, it appeared as another important feature. 
Higher the radius means weaker the bond.
It is interesting to note here the
trend of weights is the reverse than that of radius itself (Ba has the largest radius 215 pm and has the
smallest weight) for stable $BaEr_2F_8$. The scenario is reverse for unstable AuC. Ionization Energy plays similar role as Electronegativity and we observe same behavior of feature selector. As can be seen from the example, the feature selector provides elaborate cues for domain experts to reason out the results.
\begin{figure*}[h!]
	\centering
	\includegraphics[width=0.80\linewidth, height=43mm]{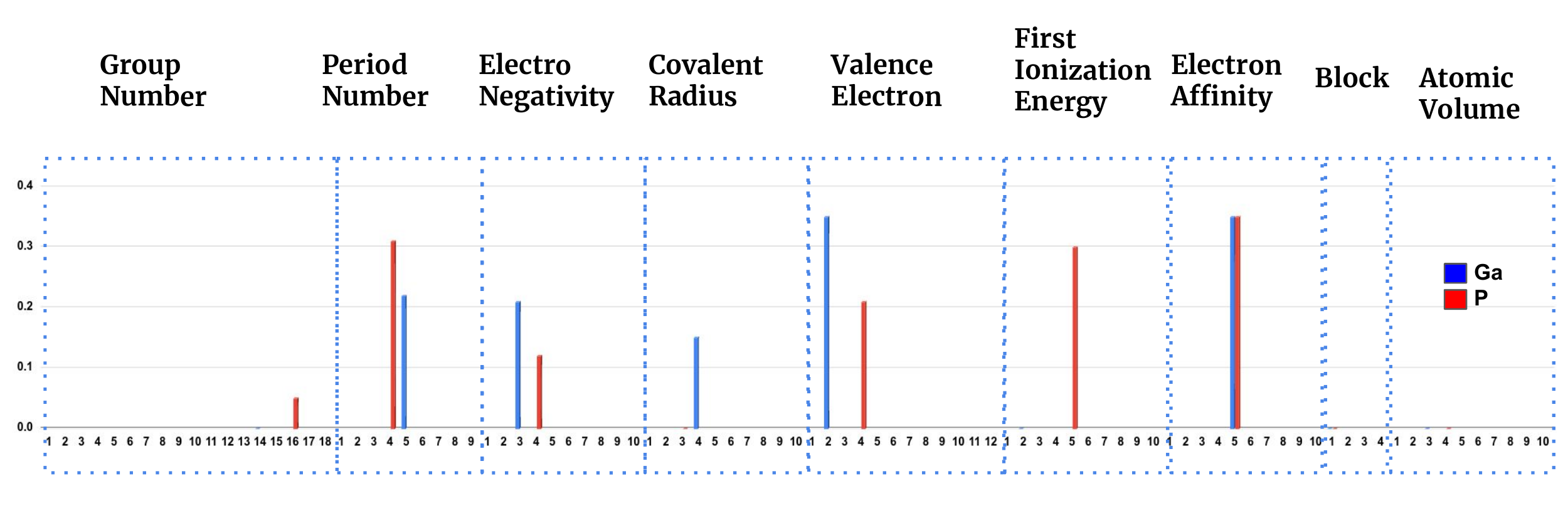}
	\caption{Feature selector values corresponding to atom features after trained on Band Gap tagged data. The top bar chart represents the feature weights of GaP (Band Gap 2.26 eV). }
	\label{fig:feature_mask_cs-g}
\end{figure*}\\
\xhdr{Band Gap}
In Fig.\ref{fig:feature_mask_cs-g}, we show the important features that appear in the case of band gap. It is interesting to see that the Electron Affinity came out to be the most
important atomic feature for the band gap as it determines the location of
conduction band minimum with respect to the vacuum.  Again, as the number of Valence Electrons and the group number are collinear properties, only one (valence electrons) is found to be having the significant weight. The Conduction Band is composed of Ga-states while the valence band is composed of P states with small admixture of Ga-states, that gives Ga-valence electrons more weight. The situation is reversed for the Ionization Energy, which determines the location of the valence band with respect to a vacuum, 
and as the valence band is mostly formed by the  P atom, we see P has  more weight.
\begin{figure*}[h!]
	\centering
	\includegraphics[width=0.80\linewidth, height=48mm]{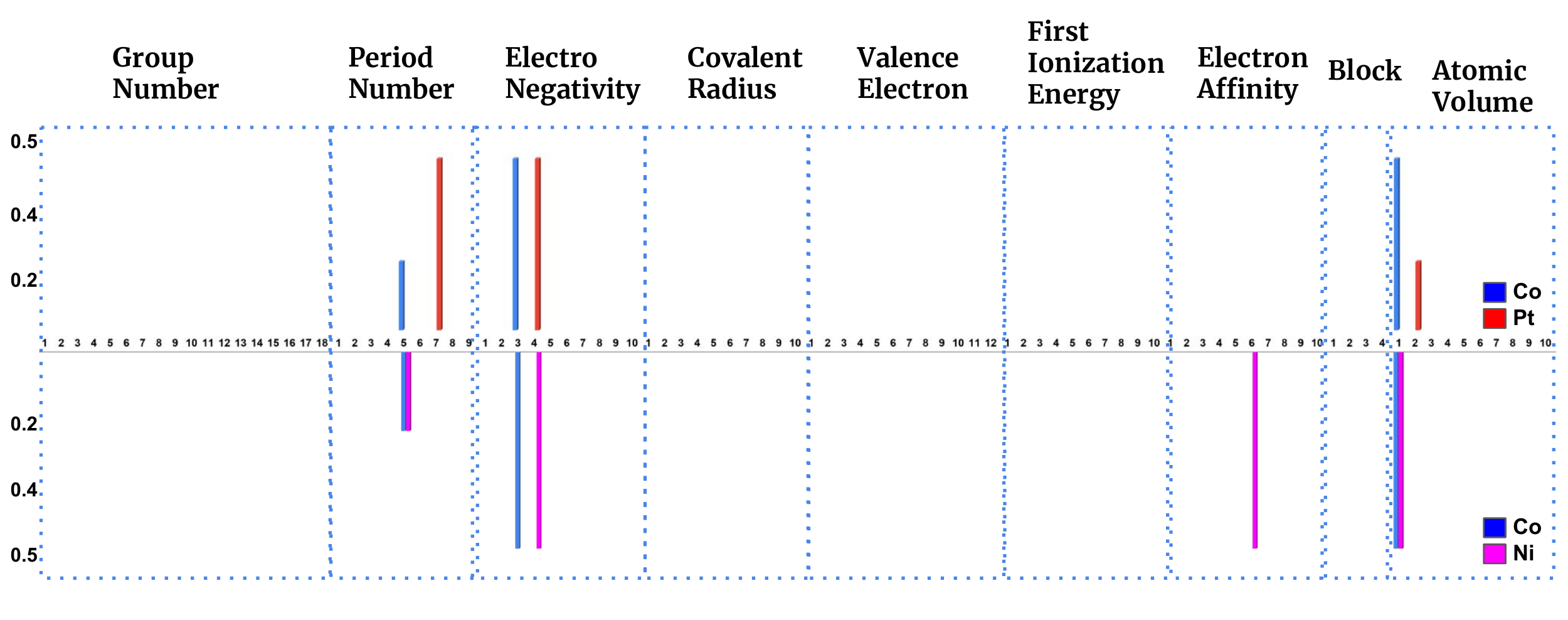}
	\caption{Feature selector values corresponding to atom features after trained on Magnetic Moment tagged data. The top bar chart represents the feature weights of CoPt and the bottom chart  represents the feature weights of CoNi. }
	\label{fig:feature_mask_cs-mm}
\end{figure*}\\
\xhdr{Magnetic Moment}
In order to understand the feature importance in the case of magnetic moment, we compare the results obtained for two Co based alloys, namely, CoPt and CoNi (Fig.\ref{fig:feature_mask_cs-mm}). In both cases  the Atomic Volume, Period Number and  Electronegativity appear to be the three most important features. While in the case of CoNi, Electron Affinity of Ni also appeared to be as additional important feature. It can be seen that in the case of CoPt, the Atomic Volume of Co has higher weight, while for CoNi, the atomic volume for both the species have the same weight. This is quite intuitive, as for CoPt, the magnetic moment is mostly carried by Co atom, while in the case of CoNi, both the atoms have significant contribution. The Electronegativity plays an important role in the context of magnetic moment. For example the magnetic moment of Co in CoPt is slightly higher than its corresponding value in pure Co. The electronegativity difference between Co and Pt causes the electron transfer from Co minority spin band to Pt which in turn enhances its magnetic moment~\cite{S1,S2}. In the case of Period Number again we see that for CoPt, it is only the period number of the magnetic atom, i.e,  the Co atom that is given visible weight while in the case of NiCo, the period number of the two atoms appears to be important.\\
It is evident from the above analysis that \our{} is effectively constructing models where the important node features are physically intuitive.\\
In conclusion, we propose an explainable property predictor for crystalline materials, \our{} to predict different crystal state and elastic properties with accurate precision using  small amount of property-tagged data. We address the issue of limited crystal data where the value of a particular property is known, using transfer learning from an encoding module CrysAE; which we train in a property agnostic way with a large amount of untagged crystal data to capture all the important structural and chemical information useful to a specific property predictor. We further find the encoder knowledge is extremely useful in de-biasing DFT error using a meagre instances of experimental results. \our{} outperforms all the baselines across seven diverse sets of properties. With appropriate case studies, we show that the explanations provided by the feature selection module  are in sync with the domain knowledge. We release the large pretrained model \ourae{} so that it could be fine-tuned using a small amount of tagged data by the research community on various applications with restricted data source.


%% file: DModel.tex
\section{Methods}
\subsection{Hyperparameters}
We have trained our model with varying convolution layers of encoder module and obtained the best results with three convolution layers in the encoder module.
We kept the embedding dimension for each node as 64, batch size of data as 512 and used average pooling to obtain $\bm{\mathcal{Z}}_g$. We selected $\lambda = 0.01$ for property selection. We varied the learning rate in logarithmic scale and selected  0.03 which yields faster convergence. We trained the auto-encoder for 200 epochs and property predictor for 200 epochs.

